\documentclass{article}

\usepackage{arxiv}

\usepackage[utf8]{inputenc} 
\usepackage[T1]{fontenc}    
\usepackage{lmodern}        
\usepackage{hyperref}       
\usepackage{url}            
\usepackage{booktabs}       
\usepackage{amsfonts}       
\usepackage{nicefrac}       
\usepackage{microtype}      
\usepackage{graphicx}

\title{sfislands: An R Package for Accommodating Islands and Disjoint Zones in Areal Spatial Modelling}

\author{
    Kevin Horan
    \thanks{Corresponding author: \href{mailto:kevin.horan.2021@mumail.ie}{\nolinkurl{kevin.horan.2021@mumail.ie}}.}
   \\
    Hamilton Institute \\
    Maynooth University \\
  Maynooth, Ireland \\
  \texttt{} \\
   \And
    Katarina Domijan
   \\
    Department of Mathematics and Statistics \\
    Maynooth University \\
  Maynooth, Ireland \\
  \texttt{} \\
   \And
    Chris Brunsdon
   \\
    National Centre for Geocomputation \\
    Maynooth University \\
  Maynooth, Ireland \\
  \texttt{} \\
  }

\usepackage{color}
\usepackage{fancyvrb}

\DefineVerbatimEnvironment{Highlighting}{Verbatim}{commandchars=\\\{\}}
\usepackage{framed}
\definecolor{shadecolor}{RGB}{248,248,248}
\newenvironment{Shaded}{\begin{snugshade}}{\end{snugshade}}

\newcommand{\AttributeTok}[1]{\textcolor[rgb]{0.13,0.29,0.53}{#1}}

\newcommand{\CommentTok}[1]{\textcolor[rgb]{0.56,0.35,0.01}{\textit{#1}}}

\newcommand{\ConstantTok}[1]{\textcolor[rgb]{0.56,0.35,0.01}{#1}}

\newcommand{\DecValTok}[1]{\textcolor[rgb]{0.00,0.00,0.81}{#1}}

\newcommand{\FloatTok}[1]{\textcolor[rgb]{0.00,0.00,0.81}{#1}}
\newcommand{\FunctionTok}[1]{\textcolor[rgb]{0.13,0.29,0.53}{\textbf{#1}}}

\newcommand{\NormalTok}[1]{#1}

\newcommand{\OtherTok}[1]{\textcolor[rgb]{0.56,0.35,0.01}{#1}}

\newcommand{\SpecialCharTok}[1]{\textcolor[rgb]{0.81,0.36,0.00}{\textbf{#1}}}

\newcommand{\StringTok}[1]{\textcolor[rgb]{0.31,0.60,0.02}{#1}}

\providecommand{\tightlist}{%
  \setlength{\itemsep}{0pt}\setlength{\parskip}{0pt}}

\usepackage{longtable,booktabs,array}
\usepackage{calc} 
\usepackage{etoolbox}
\makeatletter
\patchcmd\longtable{\par}{\if@noskipsec\mbox{}\fi\par}{}{}
\makeatother
\IfFileExists{footnotehyper.sty}{\usepackage{footnotehyper}}{\usepackage{footnote}}
\makesavenoteenv{longtable}

\NewDocumentCommand\citeproctext{}{}

\makeatletter
 \let\@cite@ofmt\@firstofone
 \def\@biblabel#1{}
 \def\@cite#1#2{{#1\if@tempswa , #2\fi}}
\makeatother
\newlength{\cslhangindent}
\newlength{\csllabelwidth}
\newenvironment{CSLReferences}[2] 
 {\begin{list}{}{%
  \setlength{\itemindent}{0pt}
  \setlength{\leftmargin}{0pt}
  \setlength{\parsep}{0pt}
  \ifodd #1
   \setlength{\leftmargin}{\cslhangindent}
   \setlength{\itemindent}{-1\cslhangindent}
  \fi
  \setlength{\itemsep}{#2\baselineskip}}}
 {\end{list}}
\usepackage{calc}

\usepackage{booktabs}
\usepackage{longtable}
\usepackage{array}
\usepackage{multirow}
\usepackage{wrapfig}
\usepackage{float}
\usepackage{colortbl}
\usepackage{pdflscape}
\usepackage{tabu}
\usepackage{threeparttable}
\usepackage{threeparttablex}
\usepackage[normalem]{ulem}
\usepackage{makecell}
\usepackage{xcolor}
\begin{document}
\maketitle

\begin{abstract}
Fitting areal models which use a spatial weights matrix to represent relationships between geographical units can be a cumbersome task, particularly when these units are not well-behaved. The two chief aims of \texttt{sfislands} are to simplify the process of creating an appropriate neighbourhood matrix, and to quickly visualise the predictions of subsequent models. The package uses visual aids in the form of easily-generated maps to help this process. This paper demonstrates how \texttt{sfislands} could be useful to researchers. It begins by describing the package's functions in the context of a proposed workflow. It then presents three worked examples showing a selection of potential use-cases. These range from earthquakes in Indonesia, to river crossings in London, and hierarchical models of output areas in Liverpool. We aim to show how the \texttt{sfislands} package streamlines much of the human workflow involved in creating and examining such models.
\end{abstract}

\keywords{
    R package
   \and
    spatial statistics
   \and
    neighbourhood structures
   \and
    ICAR models
   \and
    hierarchical spatial autoregressive modelling (HSAM)
  }

\section{Introduction}\label{introduction}

A key feature which differentiates spatial statistics is the
non-independence of observations and the expectation that neighbouring
units will be more similar than non-neighbouring ones (Tobler 1970). If this
is not accounted for, the assumptions of many types of models will be
violated. The relationships between all spatial units in a study can be
represented numerically in a spatial weights matrix. In order to build
this, we must first decide on what constitutes being a neighbour. We might
see this as a continuous relationship where degree of neighbourliness is
a function of connectivity, which could be represented as some measure
of distance. Alternatively it could be a binary situation where each
pair of units either are (1) or are not (0) neighbours. This can be
based on a condition such as contiguity of some sort, or a distance
constraint. It is the job of the modeller to formulate a hypothesis
which justifies their choice of neighbourhood structure.

For R users, the \texttt{spdep} package (R. Bivand et al. 2011) has long been popular for the
creation of these matrices. More recently, with the increasing use of \texttt{sf} structures (Pebesma 2018) and their compatibility with the \texttt{tidyverse} (Wickham et al. 2019), the \texttt{sfdep} package (Parry 2023) has presented the same
functionality in addition to extra features in a tidy structure based on
lists in a similar way to \texttt{sf}.

The most appropriate form of neighbourhood structure will depend on the specific context. Briz-Redón et al. (2021) compared different structures in the context of COVID-19 data. They note that Earnest et al. (2007) found that distance-based matrices were more appropriate when examining birth defects in Australia, whereas Duncan, White, and Mengersen (2017) found that a first-order contiguity structure produced a better fit than others in the context of lip cancer incidence in Scotland.

The most commonly used neighbourhood structure is one based on first-order queen contiguity, where units are considered neighbours if they
share at least a vortex of boundary. However, as the name suggests, this
will lead to problems when non-contiguous units such as islands or
exclaves are present. Less obviously, depending on how the geographic
units are described, areas on either sides of rivers may be
inappropriately classified as neighbours or not neighbours. Furthermore,
the presence of infrastructure such as tunnels, bridges or ferry
services might be satisfactory to meet our hypothesis of the required
degree of connectivity to be considered neighbours. Again, such information may not be apparent from a basic set of polygons. In order to create what a researcher considers to be an appropriate neighbourhood structure, incorporating all of the domain knowledge that they might have about the system, it should be simple and intuitive to add and remove connections between spatial units. This might mean adding links to account for man-made infrastructure, or cutting links to incorporate natural barriers such as rivers or mountains.

The aim of \texttt{sfislands} (Horan, Domijan, and Brunsdon 2024) is to deal with the situations described above in a
convenient and open manner. It allows us to set up a structure, quickly
map it, and then examine whether or not we are happy with how it represents our hypothesis of relationships between units. The structure can then be edited and the process re-iterated until we have described a
spatial relationship structure with which we are satisfied.

It should be noted that while this package offers convenient tools for the examination, visualisation, addition and removal of neighbourhood linkages between units, such an approach to dealing with disconnected units is not always appropriate and other methodologies are available. These issues are discussed in more depth by R. S. Bivand and Portnov (2004) and Freni-Sterrantino, Ventrucci, and Rue (2018).

The above can be considered as the \emph{pre-functions} of the package. A
second category of features, which we refer to as \emph{post-functions},
are for use after the creation of a model. Having fit a model with \texttt{mgcv} (Wood 2011) in particular, the process of extracting estimates for certain types of effects can be somewhat awkward. These \emph{post-functions} augment the original dataframe with these estimates and their standard errors in tidy format. They also allow for quick visualisation of the output in map form.

\subsection{Typical use-cases}\label{typical-use-cases}

In this paper, we will look at three examples to show different use-cases for
\texttt{sfislands}. The first example focuses on earthquakes in
Indonesia. It shows a scenario where all of the functions are used, from
setting up contiguities, to modelling and examining the predictions of the
model.

The second example looks at London and how, despite an absence of islands,
the presence of a river means that some of the pre-functions of
\texttt{sfislands} can be useful.

The final example focuses on Liverpool. There are no islands or issues
of discontiguity in this dataset. Instead, this example shows how the package can be used to quickly fit a set of reasonably complex multilevel models, with or without a spatially autocorrelated component at the lowest level, and how the results can be presented in tidy format and quickly visualised. This example also gives an opportunity to show how, when different types of contiguity from, say, the \texttt{spdep} package are used in an otherwise equivalent model, the \texttt{sfislands} post-functions make it very straightforward to quickly compare the different predictions generated from different structures.

\section{\texorpdfstring{Why use \texttt{sfislands}?}{Why use sfislands?}}\label{why-use-sfislands}

Below, we outline some of the benefits of the package in the context of a proposed workflow for fitting areal spatial models.

\subsection{\texorpdfstring{Step 1: \emph{Pre-functions} for setting up neighbourhood structure}{Step 1: Pre-functions for setting up neighbourhood structure}}\label{step-1-pre-functions-for-setting-up-neighbourhood-structure}

\begin{enumerate}
\def\labelenumi{\arabic{enumi}.}
\item
  It addresses an issue commonly seen in online help forums where an inexperienced user wishes to get started with a model but fails at the first hurdle because their neighbourhood structure contains empty records. \texttt{sfislands} will include a contiguity for all units.
\item
  It gives tools to immediately visualise this structure as a map.
\item
  These maps are created using \texttt{ggplot2} (Wickham 2016), which allows users to apply additional styling and themes using \texttt{ggplot2} syntax.
\item
  As the nodes can be labelled by index, it makes it very easy to add and remove connections as appropriate with confidence.
\item
  Connections which have been induced by a function from the package but which are not based on geographical contiguity can be accessed to ensure openness in the process.
\end{enumerate}

\subsection{Step 2: Modelling}\label{step-2-modelling}

These neighbourhood structures can be used in modelling packages such as \texttt{mgcv}, \texttt{brms} (Bürkner 2017), \texttt{r-inla} (Bakka et al. 2018) and more.

\subsection{\texorpdfstring{Step 3: \emph{Post-functions} for models}{Step 3: Post-functions for models}}\label{step-3-post-functions-for-models}

\begin{enumerate}
\def\labelenumi{\arabic{enumi}.}
\item
  It simplifies the process of extracting estimates from models, such as those with random effects and Markov random field structures created using \texttt{mgcv}. Compatibility with more packages can be added at a future date.
\item
  These effects can be quickly visualised as \texttt{ggplot2} maps.
\end{enumerate}

\section{Pre-functions}\label{pre-functions}

The first group of functions, shown in Table \ref{tab:prefunc-latex}, deals with the creation
of a neighbourhood structure in the presence of discontiguities. The
resultant structure can be quickly mapped to check if it is
satisfactory. Connections can be manually added or removed by name or
index number. By an iterative process of changes and examination of a
quickly-generated guide map, a satisfactory structure can be decided upon.

\begin{table}
\centering
\caption{\label{tab:prefunc-latex}Pre-functions: setting up a neighbourhood structure.}
\centering
\fontsize{9}{11}\selectfont
\begin{tabular}[t]{l|>{\raggedright\arraybackslash}p{7cm}}
\hline
\textbf{function} & \textbf{purpose}\\
\hline
st\_bridges() & create a neighbourhood contiguity structure, with a k-nearest neighbours condition for islands\\
\hline
st\_quickmap\_nb() & check structure visually on map\\
\hline
st\_check\_islands() & check the contiguities which have been assigned to islands\\
\hline
st\_manual\_join\_nb() & make manual changes by adding connections\\
\hline
st\_manual\_cut\_nb() & make manual changes by removing connections\\
\hline
\end{tabular}
\end{table}

\begin{figure}

{\centering \includegraphics{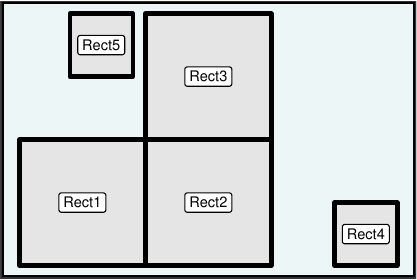} 

}

\caption{Simplified scenario with five rectangles. }\label{fig:rects1}
\end{figure}

We will now go through each function in more detail using the set of rectangles shown in Figure \ref{fig:rects1} for demonstration purposes. Rectangles 1, 2 and 3 are contiguous while 4 and 5 can be viewed as ``islands''.

\subsection{st\_bridges()}\label{st_bridges}

This function requires at least two arguments: an \texttt{sf} dataframe and, from that, the name of one column of unique identifiers, ideally names, of each spatial unit. It creates a neighbourhood structure where non-island units are joined by first-order queen contiguity, while island units are joined to their k-nearest neighbours. The output is a \emph{named} neighbourhood structure in either list or matrix form as desired, which can be either a standalone object or included as an additional column in the original \texttt{sf} dataframe. While it is not necessary in all modelling packages for the neighbourhood list or matrix to be \emph{named}, it is good practice to do so and is mandatory when using, for example, \texttt{mgcv}.

One solution when confronted with islands in a dataset is to simply exclude them from the analysis. In the first two examples of using \texttt{st\_bridges()}, we have chosen to ignore islands with the argument \texttt{remove\_islands\ =\ TRUE} and to return a list and matrix structure respectively by specifying this in the \texttt{nb\_structure} argument and choosing \texttt{add\_to\_dataframe\ =\ FALSE}:

\begin{Shaded}
\begin{Highlighting}[]
\CommentTok{\# output a named list}

\FunctionTok{st\_bridges}\NormalTok{(rectangles, }
           \StringTok{"name"}\NormalTok{, }
           \AttributeTok{remove\_islands =} \ConstantTok{TRUE}\NormalTok{, }
           \AttributeTok{nb\_structure =} \StringTok{"list"}\NormalTok{, }
           \AttributeTok{add\_to\_dataframe =} \ConstantTok{FALSE}\NormalTok{) }\SpecialCharTok{|\textgreater{}}
  \FunctionTok{head}\NormalTok{()}
\end{Highlighting}
\end{Shaded}

\begin{verbatim}
## $Rect1
## [1] 2 3
## 
## $Rect2
## [1] 1 3
## 
## $Rect3
## [1] 1 2
\end{verbatim}

\begin{Shaded}
\begin{Highlighting}[]
\CommentTok{\# output a named matrix}

\FunctionTok{st\_bridges}\NormalTok{(rectangles, }
           \StringTok{"name"}\NormalTok{, }
           \AttributeTok{remove\_islands =} \ConstantTok{TRUE}\NormalTok{, }
           \AttributeTok{nb\_structure =} \StringTok{"matrix"}\NormalTok{, }
           \AttributeTok{add\_to\_dataframe =} \ConstantTok{FALSE}\NormalTok{) }\SpecialCharTok{|\textgreater{}} 
  \FunctionTok{head}\NormalTok{()}
\end{Highlighting}
\end{Shaded}

\begin{verbatim}
##       [,1] [,2] [,3]
## Rect1    0    1    1
## Rect2    1    0    1
## Rect3    1    1    0
\end{verbatim}

Alternatively, in the following examples, we choose to join islands to their 1 nearest
neighbour, which is the default setting, and to return the output as a column called ``nb'' in the original
\texttt{sf} dataframe (\texttt{add\_to\_dataframe\ =\ "TRUE"} is the default setting):

\begin{Shaded}
\begin{Highlighting}[]
\CommentTok{\# output a named list as a column "nb" in original dataframe}

\FunctionTok{st\_bridges}\NormalTok{(rectangles, }
           \StringTok{"name"}\NormalTok{, }
           \AttributeTok{link\_islands\_k =} \DecValTok{1}\NormalTok{, }
           \AttributeTok{nb\_structure =} \StringTok{"list"}\NormalTok{) }\SpecialCharTok{|\textgreater{}} 
  \FunctionTok{head}\NormalTok{()}
\end{Highlighting}
\end{Shaded}

\begin{verbatim}
## Simple feature collection with 5 features and 2 fields
## Geometry type: POLYGON
## Dimension:     XY
## Bounding box:  xmin: 0 ymin: 0 xmax: 6 ymax: 4
## CRS:           NA
##    name      nb                       geometry
## 1 Rect1    2, 3 POLYGON ((0 0, 0 2, 2 2, 2 ...
## 2 Rect2 1, 3, 4 POLYGON ((2 0, 2 2, 4 2, 4 ...
## 3 Rect3 1, 2, 5 POLYGON ((2 2, 2 4, 4 4, 4 ...
## 4 Rect4       2 POLYGON ((5 0, 5 1, 6 1, 6 ...
## 5 Rect5       3 POLYGON ((0.8 3, 0.8 4, 1.8...
\end{verbatim}

\begin{Shaded}
\begin{Highlighting}[]
\CommentTok{\# output a named matrix as a column "nb" in original dataframe}

\FunctionTok{st\_bridges}\NormalTok{(rectangles, }
           \StringTok{"name"}\NormalTok{, }
           \AttributeTok{link\_islands\_k =} \DecValTok{1}\NormalTok{, }
           \AttributeTok{nb\_structure =} \StringTok{"matrix"}\NormalTok{) }\SpecialCharTok{|\textgreater{}} 
  \FunctionTok{head}\NormalTok{()}
\end{Highlighting}
\end{Shaded}

\begin{verbatim}
## Simple feature collection with 5 features and 2 fields
## Geometry type: POLYGON
## Dimension:     XY
## Bounding box:  xmin: 0 ymin: 0 xmax: 6 ymax: 4
## CRS:           NA
##    name nb.1 nb.2 nb.3 nb.4 nb.5                       geometry
## 1 Rect1    0    1    1    0    0 POLYGON ((0 0, 0 2, 2 2, 2 ...
## 2 Rect2    1    0    1    1    0 POLYGON ((2 0, 2 2, 4 2, 4 ...
## 3 Rect3    1    1    0    0    1 POLYGON ((2 2, 2 4, 4 4, 4 ...
## 4 Rect4    0    1    0    0    0 POLYGON ((5 0, 5 1, 6 1, 6 ...
## 5 Rect5    0    0    1    0    0 POLYGON ((0.8 3, 0.8 4, 1.8...
\end{verbatim}

These structures can serve as the input to models in \texttt{brms},
\texttt{r-inla}, \texttt{rstan} (Stan Development Team 2020) or \texttt{mgcv}. \texttt{brms} requires a matrix structure while \texttt{mgcv} models use a list. Rather than having a separate neighbours
object, it is included in the original \texttt{sf} dataframe as a named list or matrix, in the spirit of the \texttt{sfdep} package.

\subsection{st\_quickmap\_nb()}\label{st_quickmap_nb}

\begin{figure}

{\centering \includegraphics{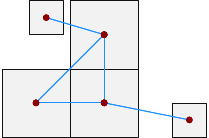} 

}

\caption{Queen contiguity and islands connected to nearest neighbour. }\label{fig:rects2}
\end{figure}

It is much more intuitive to examine these structures visually than in
matrix or list format. This can be done with the \texttt{st\_quickmap\_nb()}
function as shown in Figure \ref{fig:rects2}.

\begin{Shaded}
\begin{Highlighting}[]
\CommentTok{\# default is \textquotesingle{}nodes = "point"\textquotesingle{}}

\FunctionTok{st\_bridges}\NormalTok{(rectangles, }
           \StringTok{"name"}\NormalTok{, }
           \AttributeTok{link\_islands\_k =} \DecValTok{1}\NormalTok{) }\SpecialCharTok{|\textgreater{}} 
  \FunctionTok{st\_quickmap\_nb}\NormalTok{()}
\end{Highlighting}
\end{Shaded}

\begin{figure}

{\centering \includegraphics{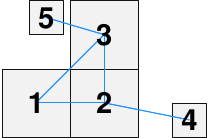} 

}

\caption{Queen contiguity and islands connected to nearest neighbour. Nodes are shown as numeric indices. }\label{fig:rects3}
\end{figure}

If we wish to make edits, it might be more useful to represent the nodes
numerically rather than as points (Figure \ref{fig:rects3}).

\begin{Shaded}
\begin{Highlighting}[]
\CommentTok{\# with \textquotesingle{}nodes = "numeric"\textquotesingle{}}

\FunctionTok{st\_bridges}\NormalTok{(rectangles, }
           \StringTok{"name"}\NormalTok{, }
           \AttributeTok{link\_islands\_k =} \DecValTok{1}\NormalTok{) }\SpecialCharTok{|\textgreater{}} 
  \FunctionTok{st\_quickmap\_nb}\NormalTok{(}\AttributeTok{nodes =} \StringTok{"numeric"}\NormalTok{)}
\end{Highlighting}
\end{Shaded}

\subsection{st\_check\_islands()}\label{st_check_islands}

This function will show us transparently what connections have been made which
are not based on contiguity. It gives both the name and index number of each pair of added connections. In this example, two pairs have been added.

\begin{Shaded}
\begin{Highlighting}[]
\CommentTok{\# show summary of non{-}contiguous connections in a dataframe}

\FunctionTok{st\_bridges}\NormalTok{(rectangles, }
           \StringTok{"name"}\NormalTok{, }
           \AttributeTok{link\_islands\_k =} \DecValTok{1}\NormalTok{) }\SpecialCharTok{|\textgreater{}} 
  \FunctionTok{st\_check\_islands}\NormalTok{()}
\end{Highlighting}
\end{Shaded}

\begin{verbatim}
##   island_names island_num nb_num nb_names
## 1        Rect4          4      2    Rect2
## 2        Rect5          5      3    Rect3
\end{verbatim}

\subsection{st\_manual\_join\_nb()}\label{st_manual_join_nb}

\begin{figure}

{\centering \includegraphics{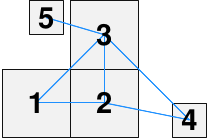} 

}

\caption{With an additional connection between 3 and 4. }\label{fig:rects4}
\end{figure}

If we feel that 4 should also be connected to 3, this can be done
manually (Figure \ref{fig:rects4}).

\begin{Shaded}
\begin{Highlighting}[]
\CommentTok{\# add an extra connection using numeric index}

\FunctionTok{st\_bridges}\NormalTok{(rectangles, }\StringTok{"name"}\NormalTok{, }
           \AttributeTok{link\_islands\_k =} \DecValTok{1}\NormalTok{) }\SpecialCharTok{|\textgreater{}} 
  \FunctionTok{st\_manual\_join\_nb}\NormalTok{(}\DecValTok{3}\NormalTok{,}\DecValTok{4}\NormalTok{) }\SpecialCharTok{|\textgreater{}} 
  \FunctionTok{st\_quickmap\_nb}\NormalTok{(}\AttributeTok{nodes =} \StringTok{"numeric"}\NormalTok{)}
\end{Highlighting}
\end{Shaded}

\subsection{st\_manual\_cut\_nb()}\label{st_manual_cut_nb}

\begin{figure}

{\centering \includegraphics{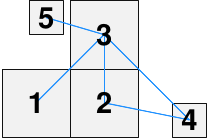} 

}

\caption{With an additional connection between 1 and 2. }\label{fig:rects5}
\end{figure}

And perhaps there is a wide river between rectangles 1 and 2 which
justifies removing the connection. We will edit it this time using
names (Figure \ref{fig:rects5}).

\begin{Shaded}
\begin{Highlighting}[]
\CommentTok{\# remove an existing connection using unit name, not index}

\FunctionTok{st\_bridges}\NormalTok{(rectangles, }\StringTok{"name"}\NormalTok{, }
           \AttributeTok{link\_islands\_k =} \DecValTok{1}\NormalTok{) }\SpecialCharTok{|\textgreater{}} 
  \FunctionTok{st\_manual\_join\_nb}\NormalTok{(}\DecValTok{3}\NormalTok{,}\DecValTok{4}\NormalTok{) }\SpecialCharTok{|\textgreater{}} 
  \FunctionTok{st\_manual\_cut\_nb}\NormalTok{(}\StringTok{"Rect1"}\NormalTok{,}\StringTok{"Rect2"}\NormalTok{) }\SpecialCharTok{|\textgreater{}} 
  \FunctionTok{st\_quickmap\_nb}\NormalTok{(}\AttributeTok{nodes =} \StringTok{"numeric"}\NormalTok{)}
\end{Highlighting}
\end{Shaded}

Having decided upon an appropriate neighbourhood structure, the next step is to use this in the context of a model. The use of such structures is particularly associated with CAR (conditional autoregressive) or ICAR-type (intrinsic conditional autoregressive) models (Besag 1974).
These are often implemented in a Bayesian framework using
\texttt{brms}, \texttt{r-inla} or \texttt{rstan}. For example, the \texttt{brms} ICAR structure requires the neighbourhood relationships to be in matrix form. The pre-functions will output the
neighbourhood structure in the desired format for use in any of these
frameworks. A convenient frequentist alternative is to use the
\texttt{mgcv} package which requires a named list of neighbours. It has the functionality to create such models
using \texttt{bs="mrf"}. It also has the ability to combine these with a
hierarchical structure using \texttt{bs="re"}. While the outputs from the
Bayesian structures mentioned above can be extracted in the same way as
any other component of the model, it can be somewhat awkward to get the
estimates from \texttt{mgcv} models. \texttt{sfislands} has two post-functions to conveniently extract and visualise these.

\section{Post-functions}\label{post-functions}

Table \ref{tab:postfunc-latex} shows the second set of functions in the package and their purpose.

\begin{table}
\centering
\caption{\label{tab:postfunc-latex}Post-functions: tidy estimates from mgcv.}
\centering
\fontsize{9}{11}\selectfont
\begin{tabular}[t]{l|l}
\hline
\textbf{function} & \textbf{purpose}\\
\hline
st\_augment() & augment the original dataframe with model predictions\\
\hline
st\_quickmap\_preds() & generate quick maps of these predictions\\
\hline
\end{tabular}
\end{table}

\subsection{st\_augment()}\label{st_augment}

This function augments the original dataframe with the estimated means
and standard errors of the spatially varying predictions from a fitted \texttt{mgcv} model in a similar manner to
how the \texttt{broom} package (Robinson, Hayes, and Couch 2023) operates. The \texttt{geometry} column, as per convention, remains as the last column of the augmented dataframe, while the predictions are positioned immediately before it. \footnote{In a similar way, \texttt{st\_augment()} can also be used to append the random effects from \texttt{lme4} (Bates et al. 2015) and \texttt{nlme} (Pinheiro, Bates, and R Core Team 2023) models to an \texttt{sf} dataframe, which can then be easily mapped using \texttt{st\_quickmap\_preds()}. Compatability with models created using different packages can be introduced in the future.} The spatially varying predictions which \texttt{st\_augment()} extracts from an \texttt{mgcv} model are

\begin{itemize}
\tightlist
\item
  random effects (which are called in \texttt{mgcv} with \texttt{bs=\textquotesingle{}re\textquotesingle{}}), and
\item
  ICAR components (\texttt{bs=\textquotesingle{}mrf\textquotesingle{}}).
\end{itemize}

Consider the model structure described in the code below using \texttt{mgcv} syntax. In this model \(y\) is the dependent variable which is being estimated with a fixed intercept, a fixed slope for some covariate, a set of random intercepts and slopes for the covariate at a \emph{region} level, and a set of ICAR varying intercepts and slopes at a lower \emph{sub-region} level.

\begin{Shaded}
\begin{Highlighting}[]
\CommentTok{\# creating an mgcv model}

\NormalTok{mgcv}\SpecialCharTok{::}\FunctionTok{gam}\NormalTok{(}
\NormalTok{  y }\SpecialCharTok{\textasciitilde{}}\NormalTok{ covariate }\SpecialCharTok{+}                   \CommentTok{\# fixed intercept and effect for covariate  }
    \FunctionTok{s}\NormalTok{(region, }\AttributeTok{bs =} \StringTok{"re"}\NormalTok{) }\SpecialCharTok{+}                  \CommentTok{\# random intercept at level region}
    \FunctionTok{s}\NormalTok{(region, covariate, }\AttributeTok{bs =} \StringTok{"re"}\NormalTok{) }\SpecialCharTok{+}          \CommentTok{\# random slopes at level region}
    \FunctionTok{s}\NormalTok{(sub}\SpecialCharTok{{-}}\NormalTok{region, }
      \AttributeTok{bs =} \StringTok{\textquotesingle{}mrf\textquotesingle{}}\NormalTok{,}
      \AttributeTok{xt =} \FunctionTok{list}\NormalTok{(}\AttributeTok{nb =}\NormalTok{ data}\SpecialCharTok{$}\NormalTok{nb), }
      \AttributeTok{k =}\NormalTok{ k) }\SpecialCharTok{+}                    \CommentTok{\# ICAR varying intercept at level sub{-}region}
    \FunctionTok{s}\NormalTok{(sub}\SpecialCharTok{{-}}\NormalTok{region, }\AttributeTok{by =}\NormalTok{ covariate, }
      \AttributeTok{bs =} \StringTok{\textquotesingle{}mrf\textquotesingle{}}\NormalTok{,}
      \AttributeTok{xt =} \FunctionTok{list}\NormalTok{(}\AttributeTok{nb =}\NormalTok{ data}\SpecialCharTok{$}\NormalTok{nb), }
      \AttributeTok{k =}\NormalTok{ k),           }\CommentTok{\# ICAR varying slope for covariate at level sub{-}region}
  \AttributeTok{data =}\NormalTok{ data, }
  \AttributeTok{method =} \StringTok{"REML"}\NormalTok{)}
\end{Highlighting}
\end{Shaded}

When labelling the new prediction columns which are augmented to the original dataframe from such a model, \texttt{st\_augment()} follows the formula syntax of the \texttt{lme4} package (Bates et al. 2015), where the pipe symbol (\texttt{\textbar{}}) indicates ``\emph{grouped by}''. Table \ref{tab:staugtab-latex} shows how the augmented columns in this scenario would be named. Each column name begins with either \texttt{random.effect.} or \texttt{mrf.smooth.} as appropriate. An additional column is also added for the standard error of each prediction, as calculated by \texttt{mgcv}. These columns are named as above but with \texttt{se.} prepended (e.g.~\texttt{se.random.effect.region}).

\begin{table}
\centering
\caption{\label{tab:staugtab-latex}The naming procedure for augmented columns from different mgcv structures.}
\centering
\fontsize{9}{11}\selectfont
\begin{tabular}[t]{l|l}
\hline
\textbf{mgcv syntax} & \textbf{column name}\\
\hline
s(region, bs = 're') & random.effect.region\\
\hline
s(region, covariate, bs = 're') & random.effect.covariate|region\\
\hline
s(sub-region, bs = 'mrf', xt = list(nb = data\$nb) & mrf.smooth.sub-region\\
\hline
s(sub-region, by = covariate, bs = 'mrf', xt = list(nb = data\$nb) & mrf.smooth.covariate|sub-region\\
\hline
\end{tabular}
\end{table}

\subsection{st\_quickmap\_preds()}\label{st_quickmap_preds}

These estimates can then be quickly mapped. As it is possible to include more than 1 spatially varying component, the output of this function is a list of plots. They can be viewed individually by indexing, or all at once using, for example, the \texttt{plotlist} argument from the \texttt{ggarrange()} function which is part of the \texttt{ggpubr} (Kassambara 2023) package. We will see this function in practice in the following examples. The maps which it generates are automatically titled and subtitled according to the type of effect. For example, the map showing predictions for \texttt{random.effect.region} will have ``\emph{region}'' as its title and ``\emph{random.effect}'' as its subtitle.

\section{Indonesia (example 1)}\label{indonesia-example-1}

Modelling earthquakes in Indonesia serves as a good example to
demonstrate this package. Firstly, Indonesia is composed of many
islands. Secondly, earthquake activity is known to be associated with
the presence of faults which exist below sea level and thus do not
respect land boundaries. Therefore it is reasonable to expect similar
behaviour in nearby provinces regardless of whether or not they are
contiguous. We aim to model the incidence, or count per unit area, of earthquake activity by
province across Indonesia, controlling for proximity to faults.

\subsection{Data}\label{data}

\begin{figure}

{\centering \includegraphics[width=1\linewidth]{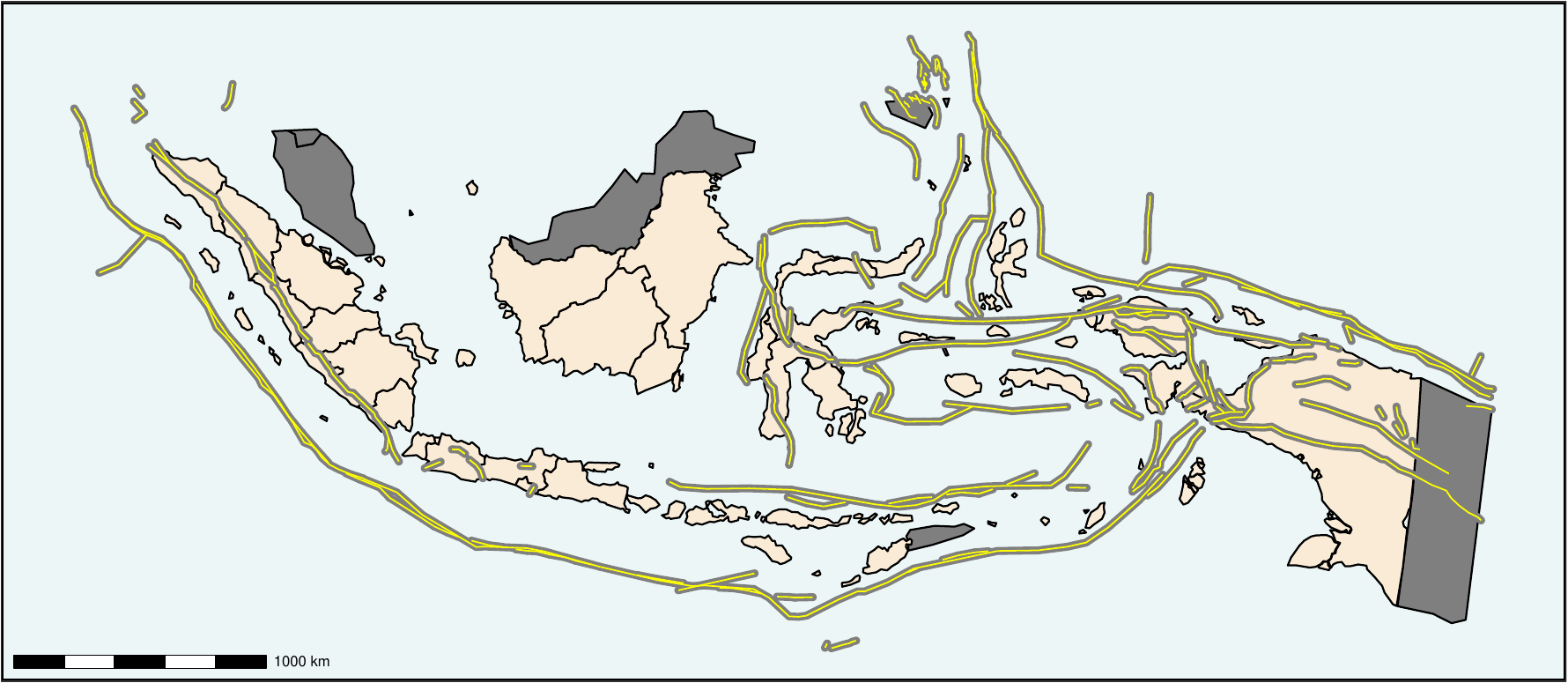} 

}

\caption{Indonesia faults. Surrounded by a 10 kilometre buffer. }\label{fig:fault-buffers}
\end{figure}

The data for this section have been downloaded from the National Earthquake Information Center, \href{https://earthquake.usgs.gov/earthquakes/search/}{USGS earthquake
catalogue}. The
datasets with accompanying explanations are available at
\url{https://github.com/horankev/quake_data}. They capture all recorded earthquakes in
and close to Indonesia from the beginning of January 1985 to the end of December 2023. Figure \ref{fig:fault-buffers} shows a map of Indonesia, divided into 33 provinces, with other neighbouring or bordering countries filled in grey. The many local
faults which lie within 300km of the shore are shown in yellow with green outlines.

\begin{figure}

{\centering \includegraphics[width=1\linewidth]{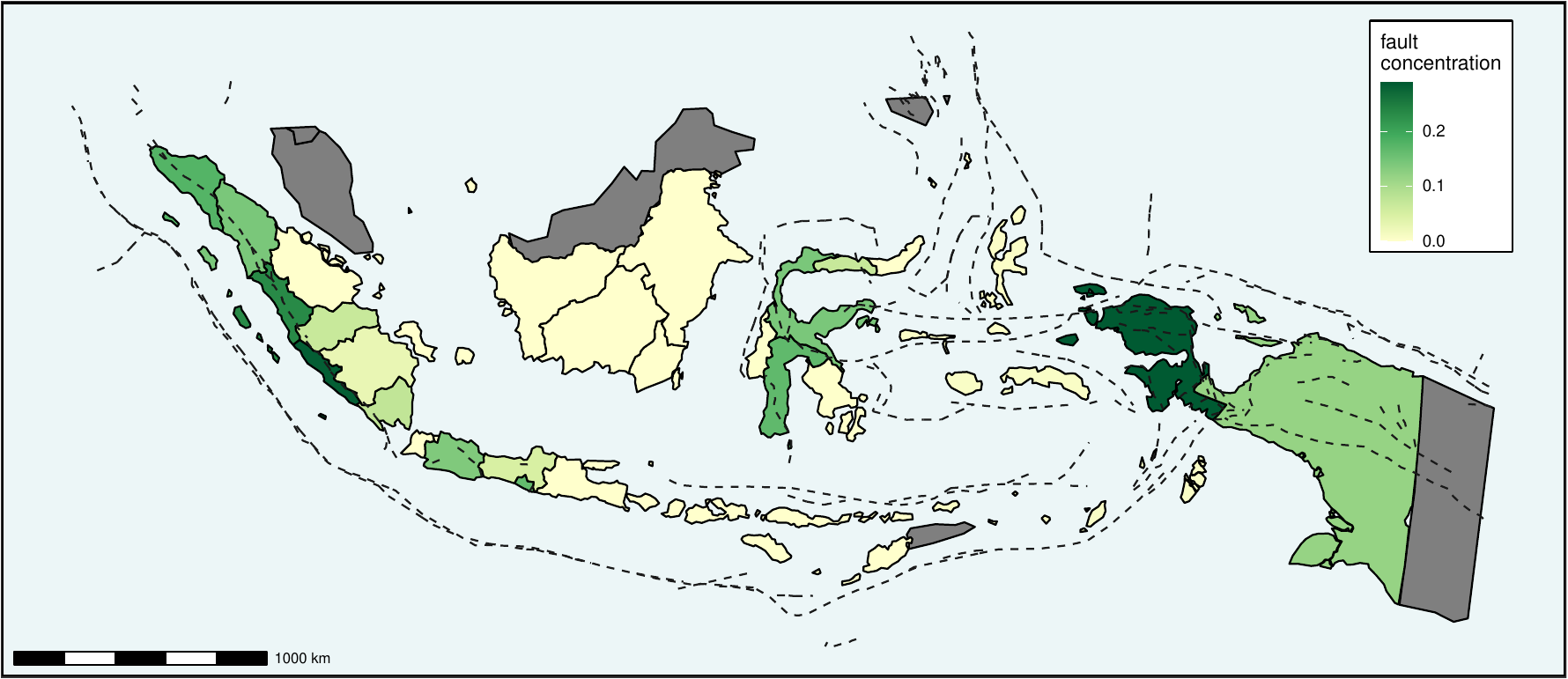} 

}

\caption{Indonesia fault concentration. Square kilometre of buffered fault per square kilometre of province area. }\label{fig:fault-conc}
\end{figure}

To get an interpretable measure of the concentration of faults in any area,
these faults are transformed from linestrings to polygons by setting a buffer of
10km around them, which explains their green outline. Now both our faults and the sizes of provinces are in units of kilometres
squared. This means we can generate a unitless metric of what proportion of any administrative unit
is covered by these buffered faults. This measure across provinces is shown in Figure
\ref{fig:fault-conc}.

Earthquake incidence per province has been calculated as the total number of earthquakes with
an epicentre within that province per unit area. We have
restricted counts to earthquakes \textgreater5.5 on the moment magnitude scale, which is the point
at which they are often labelled as potentially damaging.

\begin{figure}

{\centering \includegraphics[width=1\linewidth]{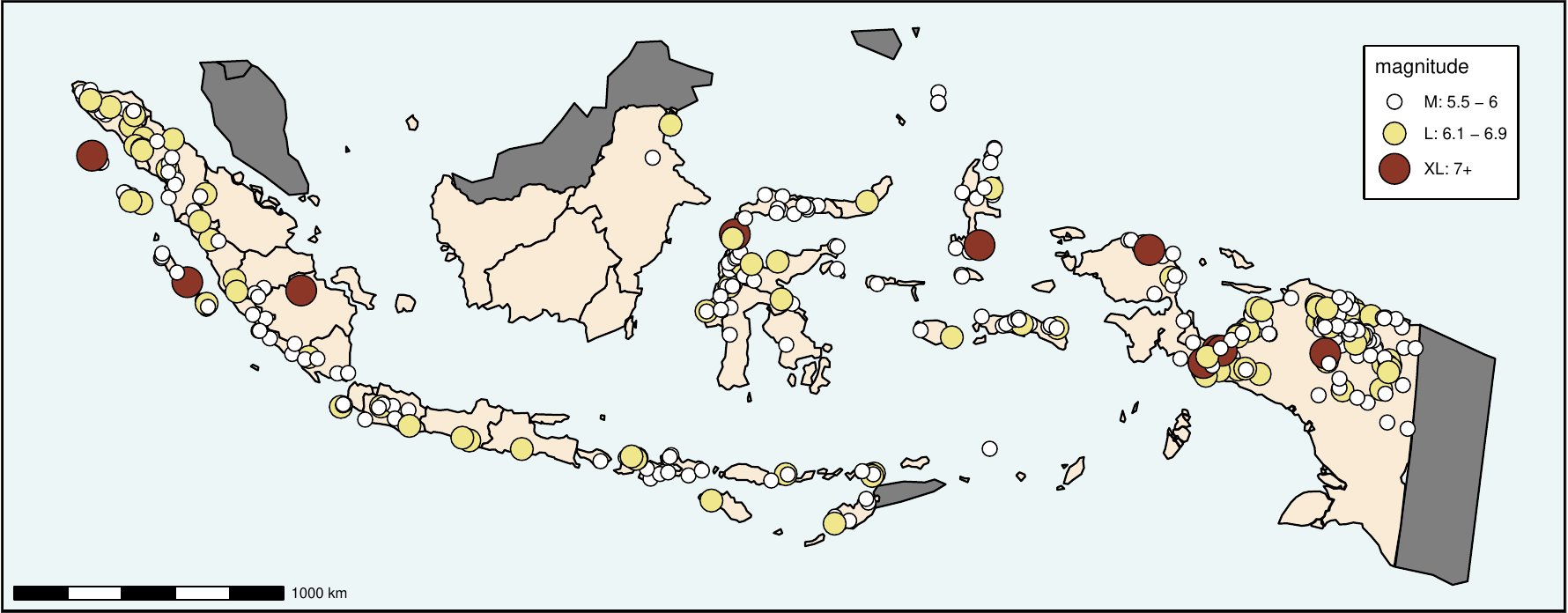} 

}

\caption{Earthquakes in Indonesia of magnitude > 5.5, 1985-2023. Categorised by magnitude as medium, large or extra-large. }\label{fig:quake-occur}
\end{figure}

\begin{figure}

{\centering \includegraphics[width=1\linewidth]{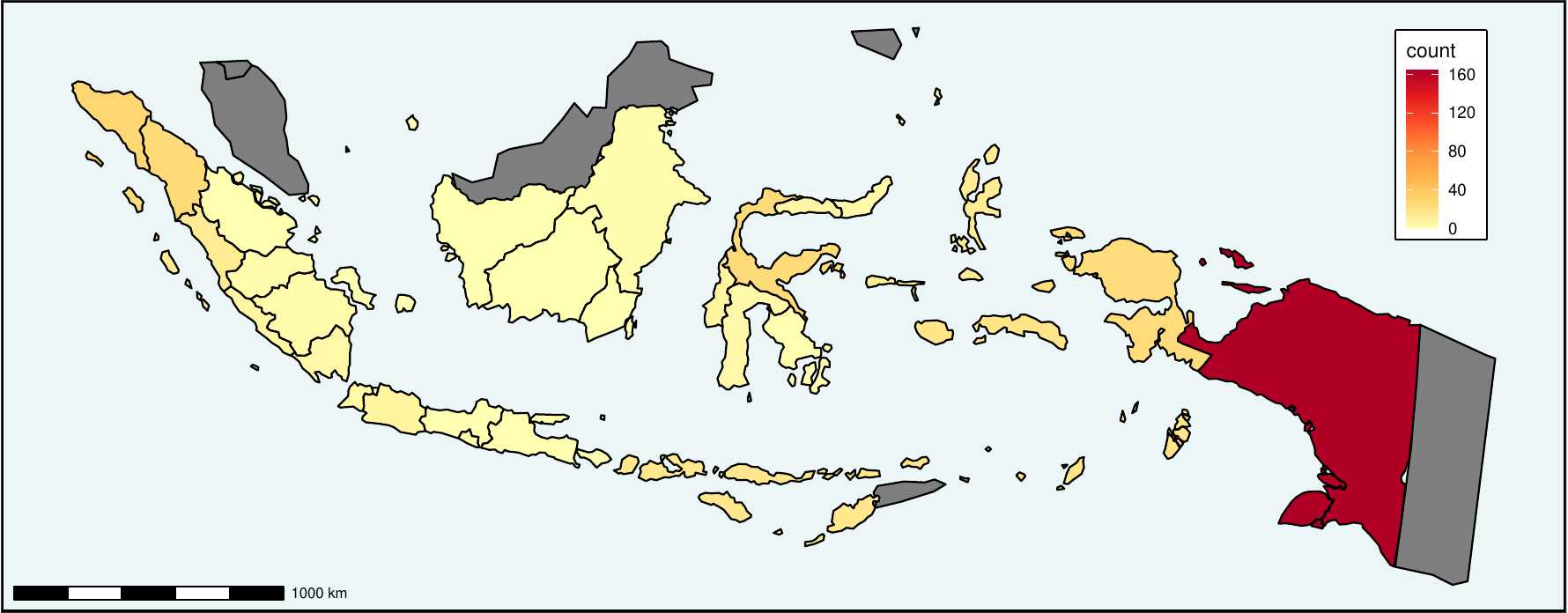} 

}

\caption{Earthquake count in Indonesia, 1985-2023, mag > 5.5: count by province. }\label{fig:quake-totals}
\end{figure}

\begin{figure}

{\centering \includegraphics[width=1\linewidth]{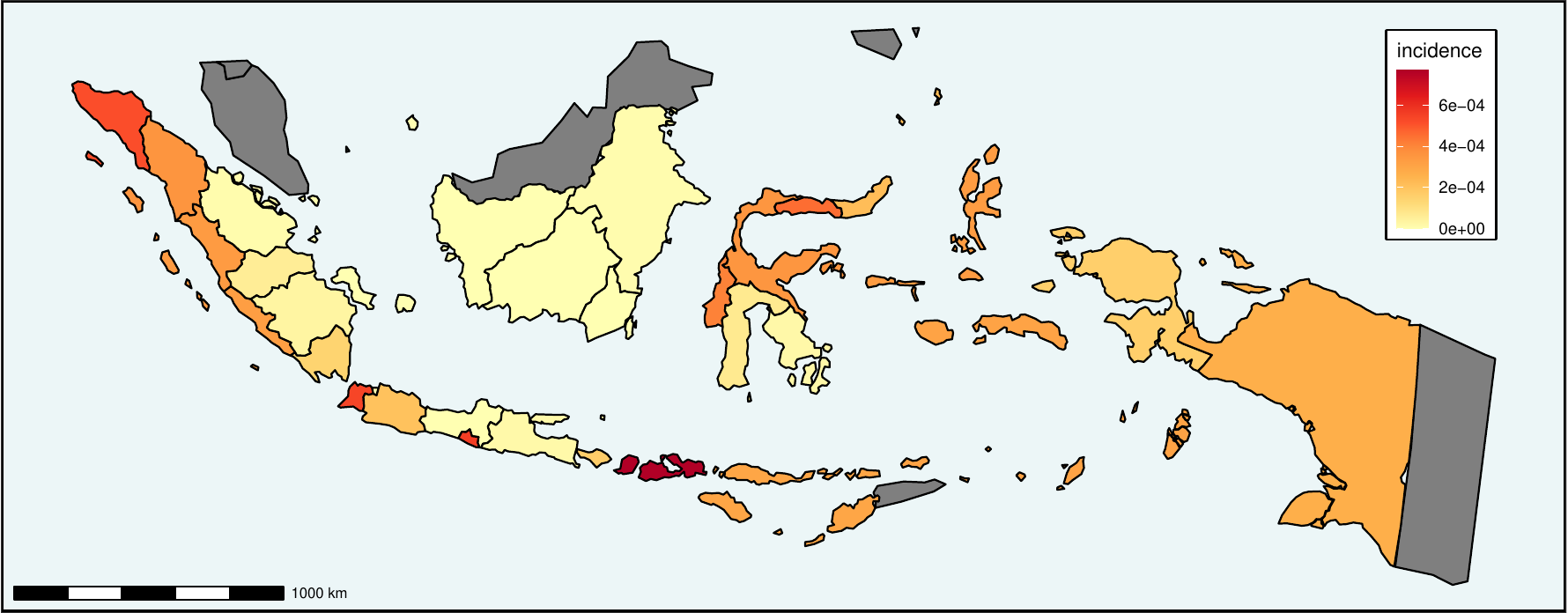} 

}

\caption{Earthquake incidence in Indonesia, 1985-2023, mag > 5.5: count per square kilometre by province. }\label{fig:quake-conc}
\end{figure}

The occurrences of these earthquakes are shown in Figure \ref{fig:quake-occur}, their total per province in Figure \ref{fig:quake-totals}, and finally, their incidence or count per square kilometre can be seen in Figure \ref{fig:quake-conc}.

\subsection{Model}\label{model}

As this is count data, we will model it as a Poisson distribution with
\(\lambda\) as the mean count per province. For \(i = 1,...,n\) provinces,
the dependent variable in this model is
\begin{equation}
y_i = \textrm{earthquake count}_i 
\label{eq:eq1}
\end{equation}
while the explanatory variable is
\begin{equation}
x_i = \textrm{fault concentration}_i = \frac{\textrm{area of buffered faults in province}_i} {\textrm{province area}_i}.
\label{eq:eq2}
\end{equation}
Firstly, when excluding the incidence and just modelling counts, where
\(y_i = \textrm{earthquake count in province}_i\), the Poisson model is of
the following form:
\begin{equation}
y_i | \lambda_i \sim \textrm{Pois}(\lambda_i)
\label{eq:eq3}
\end{equation}
with
\begin{equation}
E(y_i | \lambda_i) = \lambda_i.
\label{eq:eq4}
\end{equation}
We model
\begin{equation}
log(\lambda_i) = \beta_0 + \beta_1x_i + \gamma_i.
\label{eq:eq5}
\end{equation}
where \(\gamma_i\)
is a term with a correlation structure reflecting a province's location
relative to other provinces.

We can describe these relationships by setting up a neighbourhood
structure based on queen contiguity where a pair of provinces are
considered neighbours if they share at least one point of boundary. This can be modelled as a Markov random field to generate an ICAR model with a spatially varying term. Each of these terms will be correlated with the others according to the neighbourhood structure we have defined.

The Markov random field here follows a multivariate Gaussian
distribution. \(\gamma_i\) is a vector of province effects having a
distribution with mean \textbf{0} and precision \textbf{P} where

\([\textbf{p}]_{ij} = v_i\) if \(i=j\) and \(v_i\) is the number of adjacent provinces
to province \(i\),

\([\textbf{p}]_{ij} = -1\) if provinces \(i\) and \(j\) are adjacent, and

\([\textbf{p}]_{ij} = 0\) otherwise.

A further constraint that \(\Sigma_j \gamma_j = 0\) is applied so that the
distribution is identifiable.

We now include an offset term (here, area) because we are more interested
in modelling the incidence than in the actual count, such that
\begin{equation}
log(\frac{\lambda_i} {\textrm{area}_i}) = \beta_0 + \beta_1x_i + \gamma_i
\label{eq:eq6}
\end{equation}
which is equivalent to
\begin{equation}
log(\lambda_i) = \beta_0 + \beta_1x_i + \gamma_i + log(\textrm{area}_i).
\label{eq:eq7}
\end{equation}
We are still modelling \(log(\lambda)\) rather than the incidence, but we
are adding an offset to adjust for differing areas. Modelling \(log(\lambda)\) and adding an offset is equivalent to
modelling incidence, and coefficients can be interpreted that way.

When interpreting the estimated coefficients of the model, it can be
useful to look at it in the following form:
\begin{equation}
\lambda_i = e^{\beta_0 + \beta_1 x_i + \gamma_i} \textrm{area}_i.
\label{eq:eq8}
\end{equation}

Having described the type of model we wish to implement, we now show how \texttt{sfislands} can be used to streamline the process.

\subsection{Pre-functions}\label{pre-functions-1}

Such models, however, can not incorporate locations which have no
neighbours. In the case of Indonesia, this is quite problematic. It is
composed of many islands. The estimated count of
islands according to Andréfouët, Paul, and Farhan (2022) is 13,558.
While it is not unusual for a country to have a number of often small
offshore islands, Indonesia is entirely composed of (at least portions
of) an archipelago of islands, so many of these islands or groups of
islands are individual provinces in their own right. We might like to hypothesise that just because a province is a disconnected island, this should not mean that it is independent of other nearby provinces in terms of earthquake incidence. A standard first-order queen contiguity structure would mean the exclusion of disconnected units entirely from the model. An alternative strategy of assigning neighbour status based on a distance metric would overcome this, but the threshold size of distance necessary for such a structure might be inappropriately large for the non-islands provinces. We would like to use a compromise between these two strategies.

\begin{figure}

{\centering \includegraphics[width=1\linewidth]{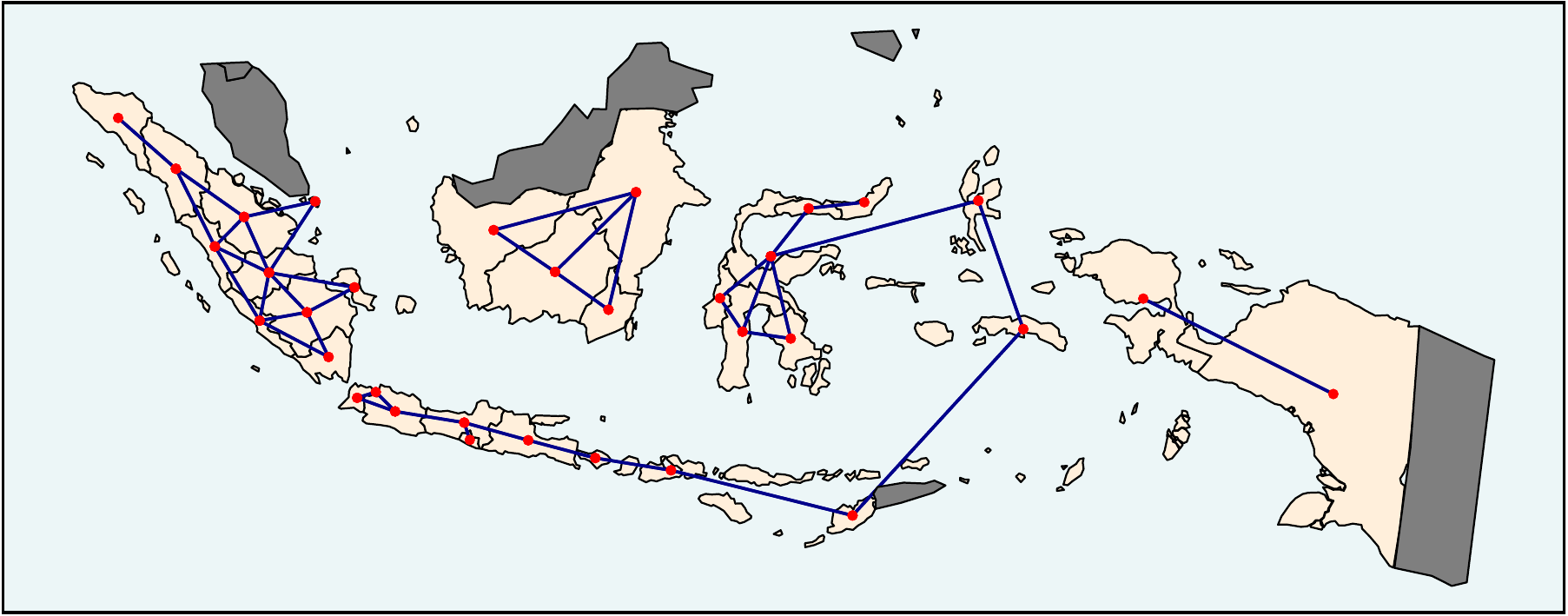} 

}

\caption{Neighbourhood structure for Indonesian provinces created by \texttt{st\_bridges()} with \texttt{k=2}.}\label{fig:ind-contig1}
\end{figure}

In this case, we use \texttt{st\_bridges()} for setting up the queen contiguity
structure as usual, but with the additional stipulation that unconnected
units (provinces which are islands or collections of islands) are
considered neighbours to their \emph{k} nearest provinces. For this example,
we have set the value of k to 2. The resulting neighbourhood structure
is shown in Figure \ref{fig:ind-contig1}. Note how is can be styled with a combination of internal arguments (size, colour, fill etc.) and additional \texttt{ggplot2} layers.

\begin{Shaded}
\begin{Highlighting}[]
\CommentTok{\# join islands to k=2 nearest neighbours}
\CommentTok{\# various arguments exist for altering colours and sizes}
\CommentTok{\# additional ggplot themes and layers can be added}

\FunctionTok{st\_bridges}\NormalTok{(provinces\_df, }\StringTok{"province"}\NormalTok{,}\AttributeTok{link\_islands\_k =} \DecValTok{2}\NormalTok{) }\SpecialCharTok{|\textgreater{}} 
  \FunctionTok{st\_quickmap\_nb}\NormalTok{(}\AttributeTok{fillcol =} \StringTok{"antiquewhite1"}\NormalTok{, }
                 \AttributeTok{bordercol =} \StringTok{"black"}\NormalTok{, }\AttributeTok{bordersize =} \FloatTok{0.5}\NormalTok{, }
                 \AttributeTok{linkcol =} \StringTok{"darkblue"}\NormalTok{, }\AttributeTok{linksize =} \FloatTok{0.8}\NormalTok{, }
                 \AttributeTok{pointcol =} \StringTok{"red"}\NormalTok{, }\AttributeTok{pointsize =} \DecValTok{2}\NormalTok{) }\SpecialCharTok{+} 
  \FunctionTok{theme}\NormalTok{(}\AttributeTok{panel.background =} \FunctionTok{element\_rect}\NormalTok{(}\AttributeTok{fill =} \StringTok{"\#ECF6F7"}\NormalTok{, }\AttributeTok{colour =} \StringTok{"black"}\NormalTok{, }
                                        \AttributeTok{linewidth=}\FloatTok{1.5}\NormalTok{),}
        \AttributeTok{axis.text =} \FunctionTok{element\_blank}\NormalTok{()) }\SpecialCharTok{+}
  \FunctionTok{geom\_sf}\NormalTok{(}\AttributeTok{data=}\NormalTok{nearby\_countries\_df, }
          \AttributeTok{fill=}\StringTok{"gray50"}\NormalTok{, }\AttributeTok{linewidth=}\FloatTok{0.5}\NormalTok{, }\AttributeTok{colour=}\StringTok{"black"}\NormalTok{)}
\end{Highlighting}
\end{Shaded}

This neighbourhood structure now has no unconnected provinces so it is
suitable for use in an ICAR model. However, if we are not entirely happy with this structure because of some domain knowledge about the inter-relationships between certain island provinces, we might wish to

\begin{itemize}
\item
  add some additional contiguities using \texttt{st\_manual\_join\_nb()}
\item
  and remove one using \texttt{st\_manual\_cut\_nb()}.
\end{itemize}

\begin{figure}

{\centering \includegraphics[width=1\linewidth]{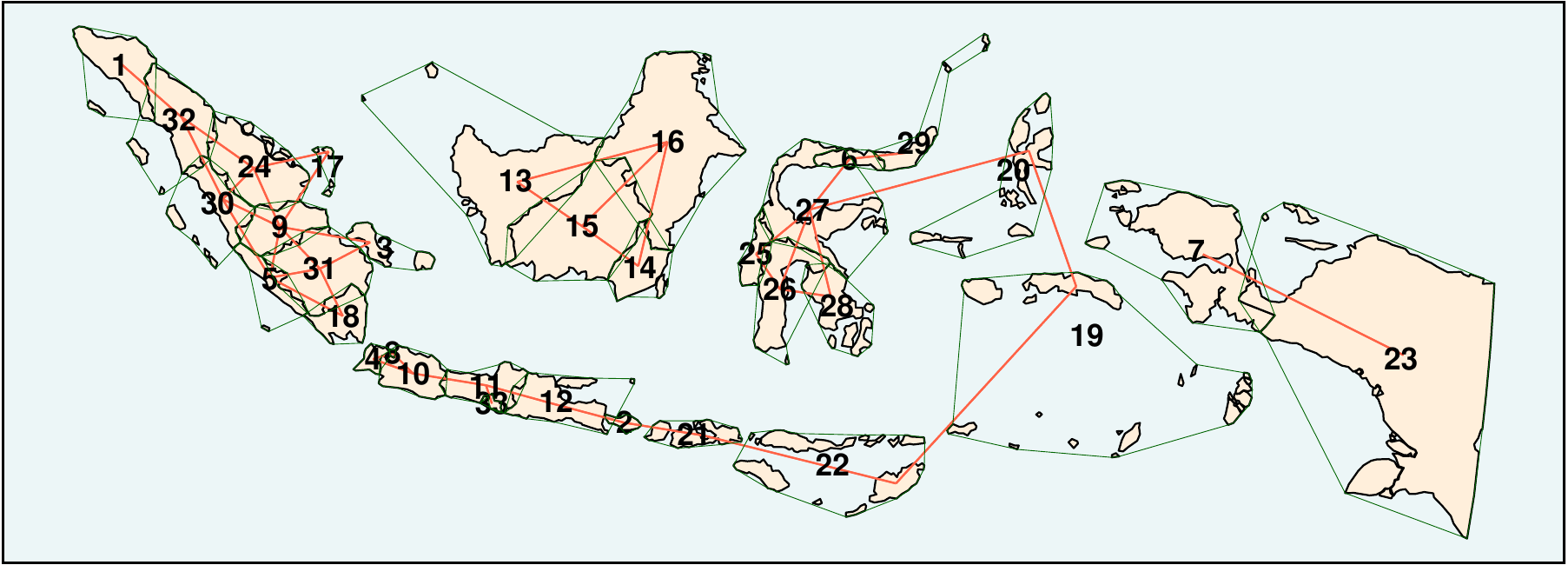} 

}

\caption{Neighbourhood structure for Indonesian provinces. Viewed with \texttt{st\_quickmap\_nb()}, using the arguments \texttt{nodes\ =\ \textquotesingle{}numeric\textquotesingle{}} and \texttt{concavehull\ =\ TRUE}.}\label{fig:ind-contig2}
\end{figure}

To cater for the possibility that a modeller might not be familiar with the names of the various geographic units but still wishes to make manual alterations to their relationships, we can look at a map (Figure \ref{fig:ind-contig2}) where the nodes are shown by
index number instead of as points (using the argument \texttt{nodes=\textquotesingle{}numeric\textquotesingle{}}). This makes it easy to manually cut
and join neighbours as desired. Furthermore, there is an option to
show concave hulls drawn around each unit (using \texttt{concavehull\ =\ TRUE}). This is also shown in Figure \ref{fig:ind-contig2}. These shapes are not used in the assignment of contiguities but it can be useful to see them in a situation such as Indonesia where many individual provinces are actually multipolygons of more than one island. Without them, it is not clear whether an island is a province in its own right, or which group of islands together form one province.

\begin{Shaded}
\begin{Highlighting}[]
\CommentTok{\# with \textquotesingle{}concavehull = TRUE\textquotesingle{} and \textquotesingle{}nodes = "numeric"\textquotesingle{}}

\FunctionTok{st\_bridges}\NormalTok{(provinces\_df, }\StringTok{"province"}\NormalTok{,}\AttributeTok{link\_islands\_k =} \DecValTok{2}\NormalTok{) }\SpecialCharTok{|\textgreater{}} 
  \FunctionTok{st\_quickmap\_nb}\NormalTok{(}\AttributeTok{fillcol =} \StringTok{"antiquewhite1"}\NormalTok{, }
                 \AttributeTok{bordercol =} \StringTok{"black"}\NormalTok{, }\AttributeTok{bordersize =} \FloatTok{0.5}\NormalTok{, }
                 \AttributeTok{linkcol =} \StringTok{"tomato"}\NormalTok{, }\AttributeTok{linksize =} \FloatTok{0.5}\NormalTok{, }
                 \AttributeTok{nodes =} \StringTok{"numeric"}\NormalTok{, }
                 \AttributeTok{numericcol =} \StringTok{"black"}\NormalTok{, }\AttributeTok{numericsize =} \DecValTok{6}\NormalTok{, }
                 \AttributeTok{concavehull =} \ConstantTok{TRUE}\NormalTok{, }
                 \AttributeTok{hullcol =} \StringTok{"darkgreen"}\NormalTok{, }\AttributeTok{hullsize =} \FloatTok{0.2}\NormalTok{) }\SpecialCharTok{+} 
  \FunctionTok{theme}\NormalTok{(}\AttributeTok{panel.background =} \FunctionTok{element\_rect}\NormalTok{(}\AttributeTok{fill =} \StringTok{"\#ECF6F7"}\NormalTok{, }\AttributeTok{colour =} \StringTok{"black"}\NormalTok{, }
                                        \AttributeTok{linewidth=}\FloatTok{1.5}\NormalTok{),}
        \AttributeTok{axis.text =} \FunctionTok{element\_blank}\NormalTok{())}
\end{Highlighting}
\end{Shaded}

\begin{figure}

{\centering \includegraphics[width=1\linewidth]{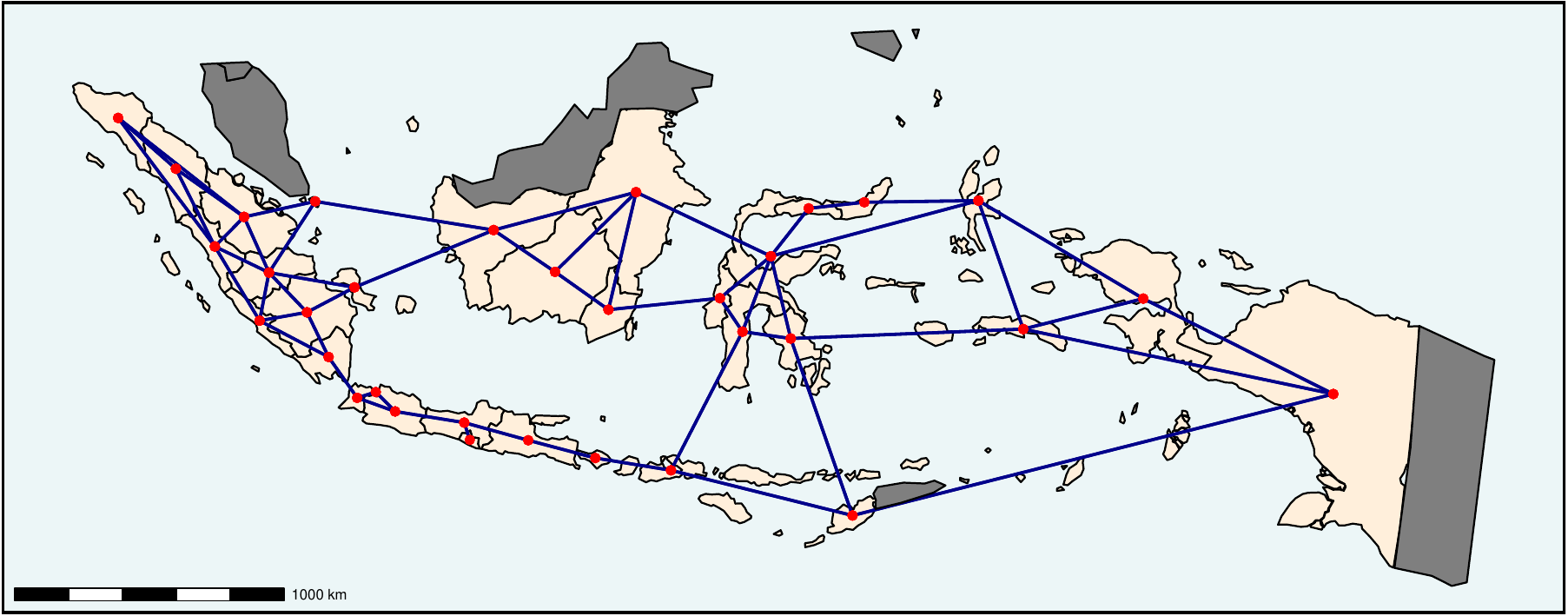} 

}

\caption{Neighbourhood structure for Indonesian provinces, after alterations using \texttt{st\_manual\_join()} and \texttt{st\_manual\_cut()}.}\label{fig:ind-contig3}
\end{figure}

Having made some manual adjustments to the neighbourhood structure, outlined in the code below, the
new structure can be seen in Figure \ref{fig:ind-contig3}. Edge effects
have also been mitigated by imposing additional connections on the two extreme
provinces (1 and 23), which would otherwise have only one neighbour, so that they now also include
their two next closest neighbours.

\begin{Shaded}
\begin{Highlighting}[]
\CommentTok{\# a series of manual joins and cuts by index number}

\FunctionTok{st\_bridges}\NormalTok{(provinces\_df, }\StringTok{"province"}\NormalTok{,}\AttributeTok{link\_islands\_k =} \DecValTok{2}\NormalTok{) }\SpecialCharTok{|\textgreater{}} 
  \FunctionTok{st\_manual\_join\_nb}\NormalTok{(}\DecValTok{1}\NormalTok{,}\DecValTok{24}\NormalTok{) }\SpecialCharTok{|\textgreater{}} 
  \FunctionTok{st\_manual\_join\_nb}\NormalTok{(}\DecValTok{1}\NormalTok{,}\DecValTok{30}\NormalTok{) }\SpecialCharTok{|\textgreater{}} 
  \FunctionTok{st\_manual\_join\_nb}\NormalTok{(}\DecValTok{3}\NormalTok{,}\DecValTok{13}\NormalTok{) }\SpecialCharTok{|\textgreater{}} 
  \FunctionTok{st\_manual\_join\_nb}\NormalTok{(}\DecValTok{13}\NormalTok{,}\DecValTok{17}\NormalTok{) }\SpecialCharTok{|\textgreater{}} 
  \FunctionTok{st\_manual\_join\_nb}\NormalTok{(}\DecValTok{14}\NormalTok{,}\DecValTok{25}\NormalTok{) }\SpecialCharTok{|\textgreater{}} 
  \FunctionTok{st\_manual\_join\_nb}\NormalTok{(}\DecValTok{20}\NormalTok{,}\DecValTok{29}\NormalTok{) }\SpecialCharTok{|\textgreater{}}
  \FunctionTok{st\_manual\_join\_nb}\NormalTok{(}\DecValTok{19}\NormalTok{,}\DecValTok{23}\NormalTok{) }\SpecialCharTok{|\textgreater{}} 
  \FunctionTok{st\_manual\_join\_nb}\NormalTok{(}\DecValTok{16}\NormalTok{,}\DecValTok{27}\NormalTok{) }\SpecialCharTok{|\textgreater{}} 
  \FunctionTok{st\_manual\_join\_nb}\NormalTok{(}\DecValTok{22}\NormalTok{,}\DecValTok{23}\NormalTok{) }\SpecialCharTok{|\textgreater{}} 
  \FunctionTok{st\_manual\_join\_nb}\NormalTok{(}\DecValTok{7}\NormalTok{,}\DecValTok{19}\NormalTok{) }\SpecialCharTok{|\textgreater{}} 
  \FunctionTok{st\_manual\_join\_nb}\NormalTok{(}\DecValTok{7}\NormalTok{,}\DecValTok{20}\NormalTok{) }\SpecialCharTok{|\textgreater{}}
  \FunctionTok{st\_manual\_join\_nb}\NormalTok{(}\DecValTok{19}\NormalTok{,}\DecValTok{28}\NormalTok{) }\SpecialCharTok{|\textgreater{}} 
  \FunctionTok{st\_manual\_join\_nb}\NormalTok{(}\DecValTok{4}\NormalTok{,}\DecValTok{18}\NormalTok{) }\SpecialCharTok{|\textgreater{}} 
  \FunctionTok{st\_manual\_join\_nb}\NormalTok{(}\DecValTok{21}\NormalTok{,}\DecValTok{26}\NormalTok{) }\SpecialCharTok{|\textgreater{}} 
  \FunctionTok{st\_manual\_join\_nb}\NormalTok{(}\DecValTok{22}\NormalTok{,}\DecValTok{28}\NormalTok{) }\SpecialCharTok{|\textgreater{}} 
  \FunctionTok{st\_manual\_cut\_nb}\NormalTok{(}\DecValTok{19}\NormalTok{,}\DecValTok{22}\NormalTok{) }\SpecialCharTok{|\textgreater{}} 
  \FunctionTok{st\_quickmap\_nb}\NormalTok{(}\AttributeTok{fillcol =} \StringTok{"antiquewhite1"}\NormalTok{, }
                 \AttributeTok{bordercol =} \StringTok{"black"}\NormalTok{, }\AttributeTok{bordersize =} \FloatTok{0.5}\NormalTok{, }
                 \AttributeTok{linkcol =} \StringTok{"darkblue"}\NormalTok{, }\AttributeTok{linksize =} \FloatTok{0.8}\NormalTok{, }
                 \AttributeTok{pointcol =} \StringTok{"red"}\NormalTok{, }\AttributeTok{pointsize =} \DecValTok{2}\NormalTok{) }\SpecialCharTok{+} 
  \FunctionTok{theme}\NormalTok{(}\AttributeTok{panel.background =} \FunctionTok{element\_rect}\NormalTok{(}\AttributeTok{fill =} \StringTok{"\#ECF6F7"}\NormalTok{, }\AttributeTok{colour =} \StringTok{"black"}\NormalTok{, }
                                        \AttributeTok{linewidth=}\FloatTok{1.5}\NormalTok{),}
        \AttributeTok{axis.text =} \FunctionTok{element\_blank}\NormalTok{()) }\SpecialCharTok{+}
  \FunctionTok{geom\_sf}\NormalTok{(}\AttributeTok{data=}\NormalTok{nearby\_countries\_df, }
          \AttributeTok{fill=}\StringTok{"gray50"}\NormalTok{, }\AttributeTok{linewidth=}\FloatTok{0.5}\NormalTok{, }\AttributeTok{colour=}\StringTok{"black"}\NormalTok{) }\SpecialCharTok{+} 
  \FunctionTok{annotation\_scale}\NormalTok{()}
\end{Highlighting}
\end{Shaded}

\subsection{\texorpdfstring{\texttt{mgcv} model}{mgcv model}}\label{mgcv-model}

We now create the ICAR model using, in this case, the \texttt{mgcv} package. We will be able to use the output of \texttt{st\_bridges}, which we have named \texttt{prep\_data}, as both the data source for the model and the neighbourhood structure (by specifying the column \texttt{nb} which contains the neighbourhood list).

\begin{Shaded}
\begin{Highlighting}[]
\NormalTok{mod\_pois\_mrf }\OtherTok{\textless{}{-}} \FunctionTok{gam}\NormalTok{(damaging\_quakes\_total }\SpecialCharTok{\textasciitilde{}} 
\NormalTok{                      fault\_concentration }\SpecialCharTok{+}
                      \FunctionTok{s}\NormalTok{(province, }\AttributeTok{bs=}\StringTok{\textquotesingle{}mrf\textquotesingle{}}\NormalTok{, }\AttributeTok{xt=}\FunctionTok{list}\NormalTok{(}\AttributeTok{nb=}\NormalTok{prep\_data}\SpecialCharTok{$}\NormalTok{nb), }\AttributeTok{k=}\DecValTok{24}\NormalTok{) }\SpecialCharTok{+}
                      \FunctionTok{offset}\NormalTok{(}\FunctionTok{log}\NormalTok{(area\_province)),}
                    \AttributeTok{data=}\NormalTok{prep\_data, }\AttributeTok{method=}\StringTok{"REML"}\NormalTok{,}\AttributeTok{family =} \StringTok{"poisson"}\NormalTok{)}
\end{Highlighting}
\end{Shaded}

We can see from the summary below that the adjusted R-squared is \textbf{0.983} and deviance explained is \textbf{93.3\%}. The coefficient for \texttt{fault\_concentration} confirms an expected positive mean global association between earthquake and fault incidence.

\begin{verbatim}
## 
## Family: poisson 
## Link function: log 
## 
## Formula:
## damaging_quakes_total ~ fault_concentration + s(province, bs = "mrf", 
##     xt = list(nb = prep_data$nb), k = 24) + offset(log(area_province))
## 
## Parametric coefficients:
##                     Estimate Std. Error z value Pr(>|z|)    
## (Intercept)          -9.5648     0.1744 -54.845  < 2e-16 ***
## fault_concentration   5.9971     1.9245   3.116  0.00183 ** 
## ---
## Signif. codes:  0 '***' 0.001 '**' 0.01 '*' 0.05 '.' 0.1 ' ' 1
## 
## Approximate significance of smooth terms:
##               edf Ref.df Chi.sq p-value    
## s(province) 19.19     23  166.6  <2e-16 ***
## ---
## Signif. codes:  0 '***' 0.001 '**' 0.01 '*' 0.05 '.' 0.1 ' ' 1
## 
## R-sq.(adj) =  0.983   Deviance explained = 93.3%
## -REML = 104.81  Scale est. = 1         n = 33
\end{verbatim}

Returning to the initial question, what is the additional risk level of earthquakes in a province, having controlled for the concentration of faults? This can be seen as a measure of the activity level of faults locally and it is spatially smoothed by the autoregressive process. It is represented in the model summary by the component \texttt{s(province)}. However, the extraction of individual predictions for this component for each province from the \texttt{mgcv} model requires a number of steps. We now demonstrate how these are streamlined into a single function by the \texttt{sfislands} package.

\subsection{Post-functions}\label{post-functions-1}

The function \texttt{st\_augment()} allow us to add the spatially varying predictions from the model as new columns to the original dataframe in a process
similar to that of the \texttt{broom} package. For instance, we see from the output of the following code chunk that the original dataframe is now augmented with columns called \texttt{mrf.smooth.province} and \texttt{se.mrf.smooth.province} which show the predictions for the \(\gamma_i\) component and their standard errors. Note that this is how we would expect them to be named, based on the previous discussion surrounding Table \ref{tab:staugtab-latex}. They are positioned immediately before the final \texttt{geometry} column of the \texttt{sf} dataframe, and after the neighbours list column, \texttt{nb}.

\begin{Shaded}
\begin{Highlighting}[]
\CommentTok{\# column names of augmented dataframe}

\NormalTok{mod\_pois\_mrf }\SpecialCharTok{|\textgreater{}} 
  \FunctionTok{st\_augment}\NormalTok{(prep\_data) }\SpecialCharTok{|\textgreater{}} 
  \FunctionTok{names}\NormalTok{()}
\end{Highlighting}
\end{Shaded}

\begin{verbatim}
##  [1] "province"                "province_id"            
##  [3] "S"                       "M"                      
##  [5] "L"                       "XL"                     
##  [7] "quake_total"             "quake_density"          
##  [9] "damaging_quakes_total"   "damaging_quakes_density"
## [11] "area_fault_within"       "area_province"          
## [13] "fault_concentration"     "nb"                     
## [15] "mrf.smooth.province"     "se.mrf.smooth.province" 
## [17] "geometry"
\end{verbatim}

\begin{figure}

{\centering \includegraphics[width=1\linewidth]{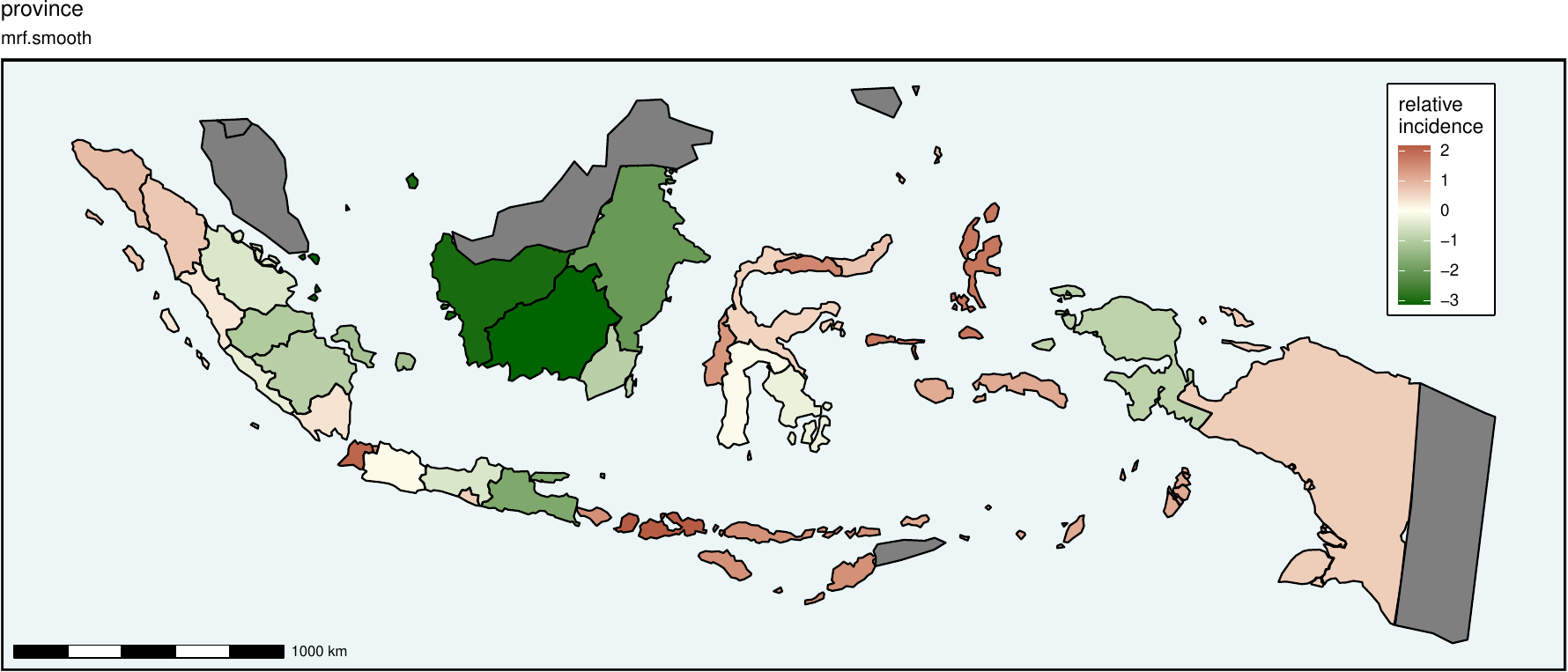} 

}

\caption{Estimates of \(\gamma_i\) shown as a map using \texttt{st\_quickmap\_preds()}.}\label{fig:ind-mrf1}
\end{figure}

This output can now be piped into the \texttt{st\_quickmap\_preds()} function to get a quick visualisation of
these estimates for \(\gamma_i\) on a map, as shown in Figure \ref{fig:ind-mrf1}. Again, note that the title and subtitle of the image are as previously discussed.

\begin{Shaded}
\begin{Highlighting}[]
\CommentTok{\# st\_quickmap\_preds() outputs a list of ggplots}

\NormalTok{plot\_mrf }\OtherTok{\textless{}{-}}\NormalTok{ mod\_pois\_mrf }\SpecialCharTok{|\textgreater{}} 
  \FunctionTok{st\_augment}\NormalTok{(prep\_data) }\SpecialCharTok{|\textgreater{}}
  \FunctionTok{st\_quickmap\_preds}\NormalTok{(}\AttributeTok{scale\_low =} \StringTok{"darkgreen"}\NormalTok{,}
                    \AttributeTok{scale\_mid =} \StringTok{"ivory"}\NormalTok{, }
                    \AttributeTok{scale\_high =} \StringTok{"darkred"}\NormalTok{, }
                    \AttributeTok{scale\_midpoint =} \DecValTok{0}\NormalTok{)}

\CommentTok{\# in this case, there is only one plot in the list}
\CommentTok{\# so we call it by index}
\CommentTok{\# it is then supplemented with additional ggplot functions}

\NormalTok{plot\_mrf[[}\DecValTok{1}\NormalTok{]] }\SpecialCharTok{+}
  \FunctionTok{coord\_sf}\NormalTok{(}\AttributeTok{datum=}\ConstantTok{NA}\NormalTok{) }\SpecialCharTok{+}
  \FunctionTok{theme}\NormalTok{(}\AttributeTok{panel.background =} \FunctionTok{element\_rect}\NormalTok{(}\AttributeTok{fill =} \StringTok{"\#ECF6F7"}\NormalTok{, }\AttributeTok{colour =} \StringTok{"black"}\NormalTok{, }
                                        \AttributeTok{linewidth=}\FloatTok{1.5}\NormalTok{),}
        \AttributeTok{axis.text =} \FunctionTok{element\_blank}\NormalTok{()) }\SpecialCharTok{+}
  \FunctionTok{geom\_sf}\NormalTok{(}\AttributeTok{data=}\NormalTok{provinces\_df, }\AttributeTok{fill=}\ConstantTok{NA}\NormalTok{, }\AttributeTok{colour=}\StringTok{"black"}\NormalTok{, }\AttributeTok{linewidth=}\FloatTok{0.5}\NormalTok{) }\SpecialCharTok{+} 
  \FunctionTok{geom\_sf}\NormalTok{(}\AttributeTok{data=}\NormalTok{nearby\_countries\_df, }\AttributeTok{fill=}\StringTok{"gray50"}\NormalTok{, }\AttributeTok{colour=}\StringTok{"black"}\NormalTok{, }
          \AttributeTok{linewidth=}\FloatTok{0.5}\NormalTok{) }\SpecialCharTok{+} 
  \FunctionTok{labs}\NormalTok{(}\AttributeTok{fill=}\StringTok{"relative}\SpecialCharTok{\textbackslash{}n}\StringTok{incidence"}\NormalTok{) }\SpecialCharTok{+}
  \FunctionTok{annotation\_scale}\NormalTok{() }\SpecialCharTok{+}
  \FunctionTok{coord\_sf}\NormalTok{(}\AttributeTok{datum=}\ConstantTok{NA}\NormalTok{) }\SpecialCharTok{+} 
  \FunctionTok{theme}\NormalTok{(}\AttributeTok{legend.position =} \FunctionTok{c}\NormalTok{(}\FloatTok{0.92}\NormalTok{,}\FloatTok{0.77}\NormalTok{),}
        \AttributeTok{legend.box.background =} \FunctionTok{element\_rect}\NormalTok{(}\AttributeTok{colour =} \StringTok{"black"}\NormalTok{, }\AttributeTok{linewidth =} \DecValTok{1}\NormalTok{),}
        \AttributeTok{legend.title =} \FunctionTok{element\_text}\NormalTok{())}
\end{Highlighting}
\end{Shaded}

\begin{figure}

{\centering \includegraphics[width=1\linewidth]{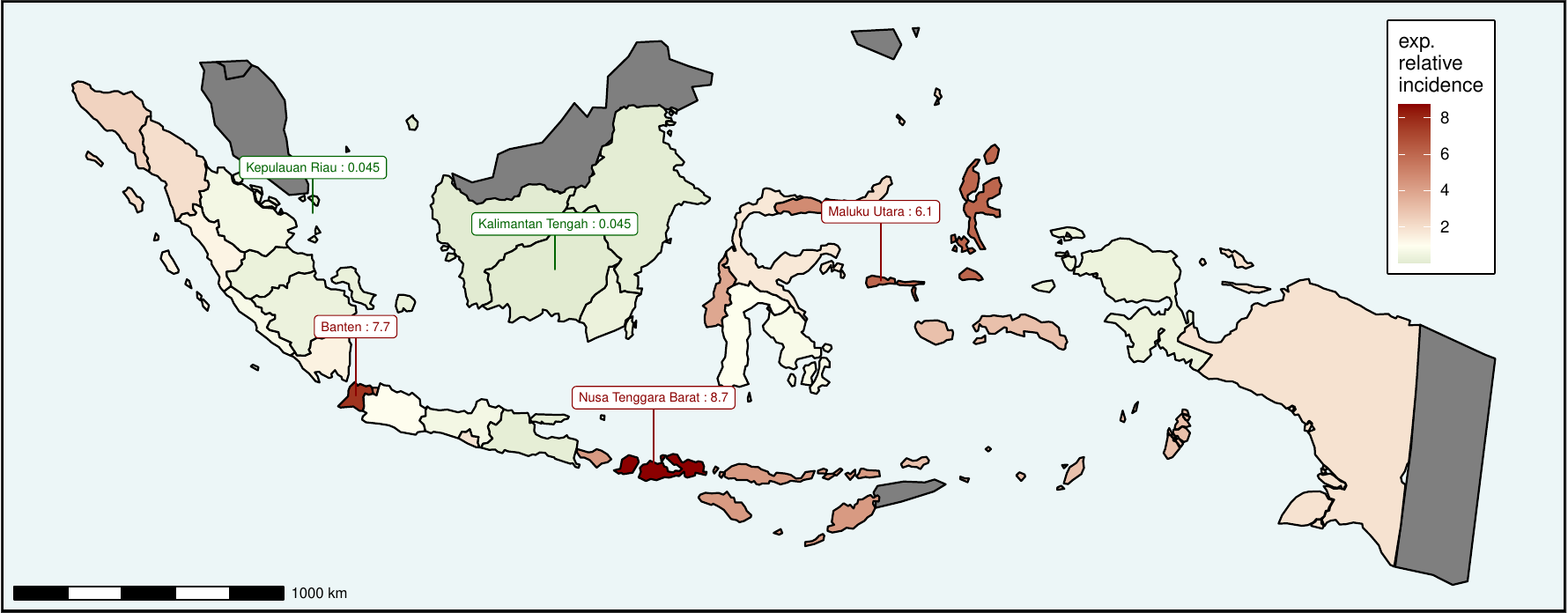} 

}

\caption{Map showing estimates of \(\textrm{exp}(\gamma_i).\) This is produced by adding an additional column to the dataframe produced by \texttt{st\_augment()}.}\label{fig:ind-mrf2}
\end{figure}

If we wish to apply the inverse link function (the exponential function in the case of this Poisson model) to map these values to a more interpretable scale, this will not be generated by the function \texttt{st\_quickmap\_preds()}. Instead, we must use the augmented dataframe which is produced by \texttt{st\_augment()} and create the appropriate extra column with the usual \texttt{tidyverse} \texttt{mutate()} function. This allows us to produce the map in Figure
\ref{fig:ind-mrf2}. As these coefficients are multiplicatively related to the earthquake incidence, values below 1 imply an earthquake incidence which is lower than expected.

The provinces with the 3 most elevated incidences are labelled in red. We can see that, controlling for the effects of proximity to faults, the province of Nusa Tenggara Barat has 8.7 times the expected incidence, or number of major earthquakes per square kilometre. The two lowest-scoring provinces, labelled in green, have essentially no incidence of earthquake epicentres within their boundaries, controlling for what their proximity to faults alone would suggest.

\subsection{Workflow summary}\label{workflow-summary}

In this example, we have gone through a number of stages carefully, making changes to contiguities that we deemed appropriate as we went. However, in practice, at least in a first iteration, it might not be necessary to go through all of these steps. A rough and ready model, complete with spatially varying coefficients and visual output, can be generated with \texttt{sfislands} using nothing more than three or four lines of code, such as the following:

\begin{Shaded}
\begin{Highlighting}[]
\CommentTok{\# workflow:}

\CommentTok{\# 1. set up neighbourhood structure}

\NormalTok{prep\_data }\OtherTok{\textless{}{-}} \FunctionTok{st\_bridges}\NormalTok{(provinces\_df, }\StringTok{"province"}\NormalTok{)}

\CommentTok{\# 2. define model}

\NormalTok{mod }\OtherTok{\textless{}{-}} \FunctionTok{gam}\NormalTok{(quake\_mlxl\_total }\SpecialCharTok{\textasciitilde{}} 
\NormalTok{                  fault\_concentration }\SpecialCharTok{+}
                  \FunctionTok{s}\NormalTok{(province, }\AttributeTok{bs=}\StringTok{\textquotesingle{}mrf\textquotesingle{}}\NormalTok{, }\AttributeTok{xt=}\FunctionTok{list}\NormalTok{(}\AttributeTok{nb=}\NormalTok{prep\_data}\SpecialCharTok{$}\NormalTok{nb), }\AttributeTok{k=}\DecValTok{22}\NormalTok{) }\SpecialCharTok{+}
                  \FunctionTok{offset}\NormalTok{(}\FunctionTok{log}\NormalTok{(area\_province)),}
        \AttributeTok{data=}\NormalTok{prep\_data, }\AttributeTok{method=}\StringTok{"REML"}\NormalTok{,}\AttributeTok{family =} \StringTok{"poisson"}\NormalTok{)}

\CommentTok{\# 3. augment tidy estimates}

\NormalTok{tidy\_ests }\OtherTok{\textless{}{-}} \FunctionTok{st\_augment}\NormalTok{(mod, prep\_data)}

\CommentTok{\# 4. visualise them}

\FunctionTok{st\_quickmap\_preds}\NormalTok{(tidy\_ests)}
\end{Highlighting}
\end{Shaded}

\section{London (example 2)}\label{london-example-2}

\begin{figure}

{\centering \includegraphics[width=1\linewidth]{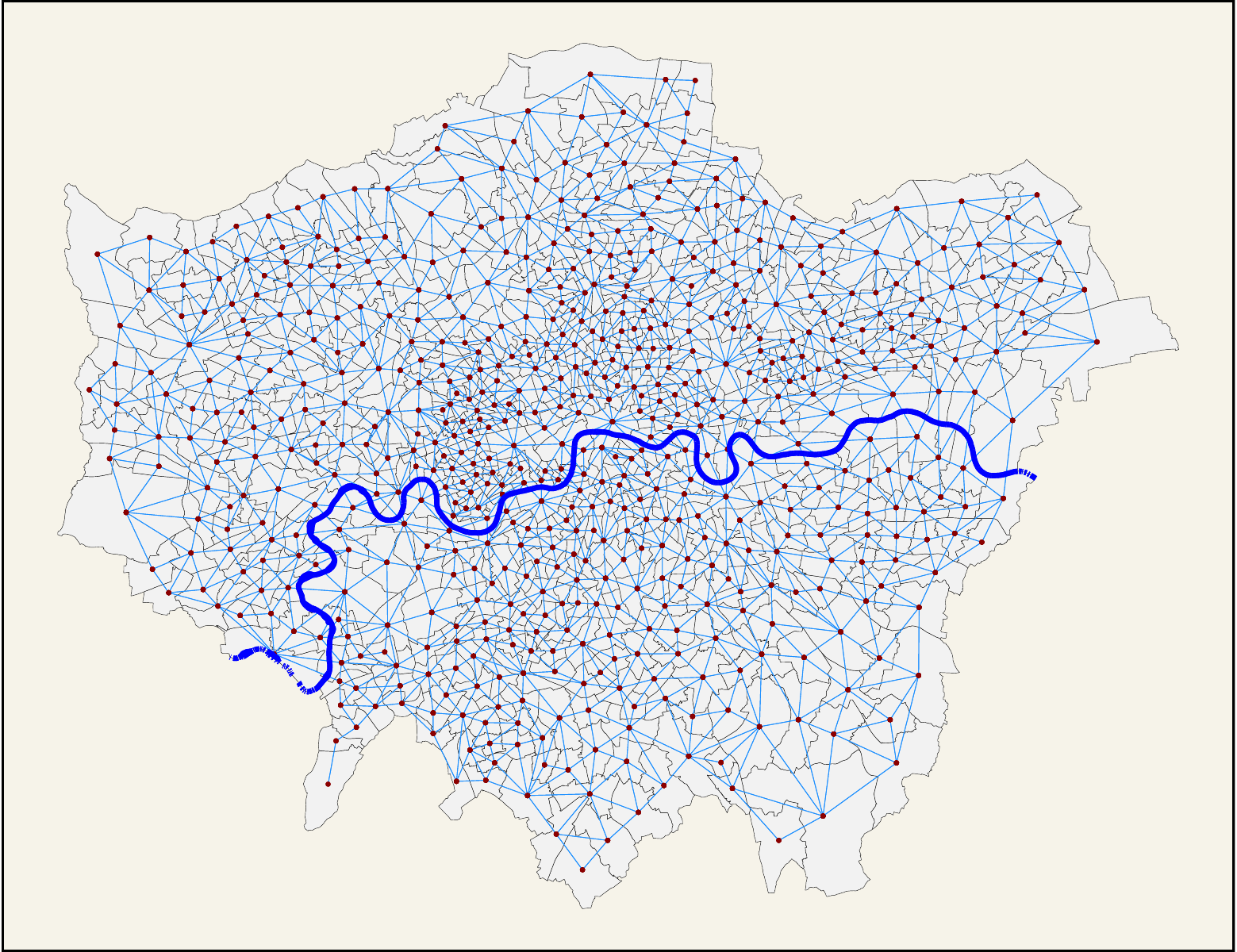} 

}

\caption{Wards and boroughs of Greater London. Queen contiguity. }\label{fig:lon-contig1}
\end{figure}

The next example looks only at using the \emph{pre-functions} of \texttt{sfislands}, but in a situation where the presence of actual \emph{islands} is not the problem we seek to deal with. Consider the wards and boroughs of London (sourced from the Greater London Authority's \href{https://data.london.gov.uk/}{London Datastore}) and available at
\url{https://github.com/horankev/london_liverpool_data}. In Figure \ref{fig:lon-contig1} the \texttt{st\_bridges()}
function is applied to them to construct a queen contiguity
neighbourhood structure. Because there are no disjoint units (or
``islands''), this will be the same as using \texttt{st\_contiguity()} from
\texttt{sfdep}. The \texttt{st\_quickmap\_nb()} function gives an immediate
visual representation of the structure. \footnote{\texttt{st\_quickmap\_nb()} can also be used to visualise any contiguity structure created by \texttt{spdep} or \texttt{sfdep} as long as that structure is included in an \texttt{sf} dataframe as a column named \texttt{nb}.} Because this map is created using \texttt{ggplot2}, it can be easily supplemented by adding a layer showing the course of the river Thames which is also visible in Figure \ref{fig:lon-contig1}.

\begin{Shaded}
\begin{Highlighting}[]
\CommentTok{\# same as sfdep:st\_contiguity() as there are no islands}
\CommentTok{\# an extra layer for the river Thames}

\FunctionTok{st\_bridges}\NormalTok{(london, }\StringTok{"NAME"}\NormalTok{) }\SpecialCharTok{|\textgreater{}} 
  \FunctionTok{st\_quickmap\_nb}\NormalTok{() }\SpecialCharTok{+} 
  \FunctionTok{geom\_sf}\NormalTok{(}\AttributeTok{data=}\NormalTok{thames, }\AttributeTok{colour=}\StringTok{"blue"}\NormalTok{, }\AttributeTok{linewidth=}\FloatTok{1.5}\NormalTok{) }\SpecialCharTok{+} 
  \FunctionTok{theme}\NormalTok{(}\AttributeTok{panel.background =} \FunctionTok{element\_rect}\NormalTok{(}\AttributeTok{fill =} \StringTok{"\#F6F3E9"}\NormalTok{, }\AttributeTok{colour =} \StringTok{"black"}\NormalTok{, }
                                        \AttributeTok{linewidth=}\FloatTok{1.5}\NormalTok{))}
\end{Highlighting}
\end{Shaded}

When a study area has a river running through it, problems can arise with constructing appropriate neighbourhood structures. Depending on how the
geometries are defined, the presence of a river can cause problems in two ways. In one situation, the river could be expressed as a polygon in its own right meaning that, using the condition of queen contiguity, it severs any potential contiguity between units on either side of its banks. In this situation, no spatial units will be neighbours with the units directly across the river from them. At the other extreme, if the river is not included as a geometry (as is the case here) all units on opposing banks are automatically considered neighbours.

Depending on the presence of river crossings, two areas which are physically quite close but on opposing banks might be very distinct. If there is no means of crossing the river within a reasonable distance, somebody living on the banks of a river might be more likely to go about their life primarily on their side of the river, despite the short distance as the crow flies of facilities on the other side. This could be relevant in terms of, say, modelling of house prices where we might want to incorporate issues such as local amenities into a neighbourhood structure.

\begin{figure}

{\centering \includegraphics[width=1\linewidth]{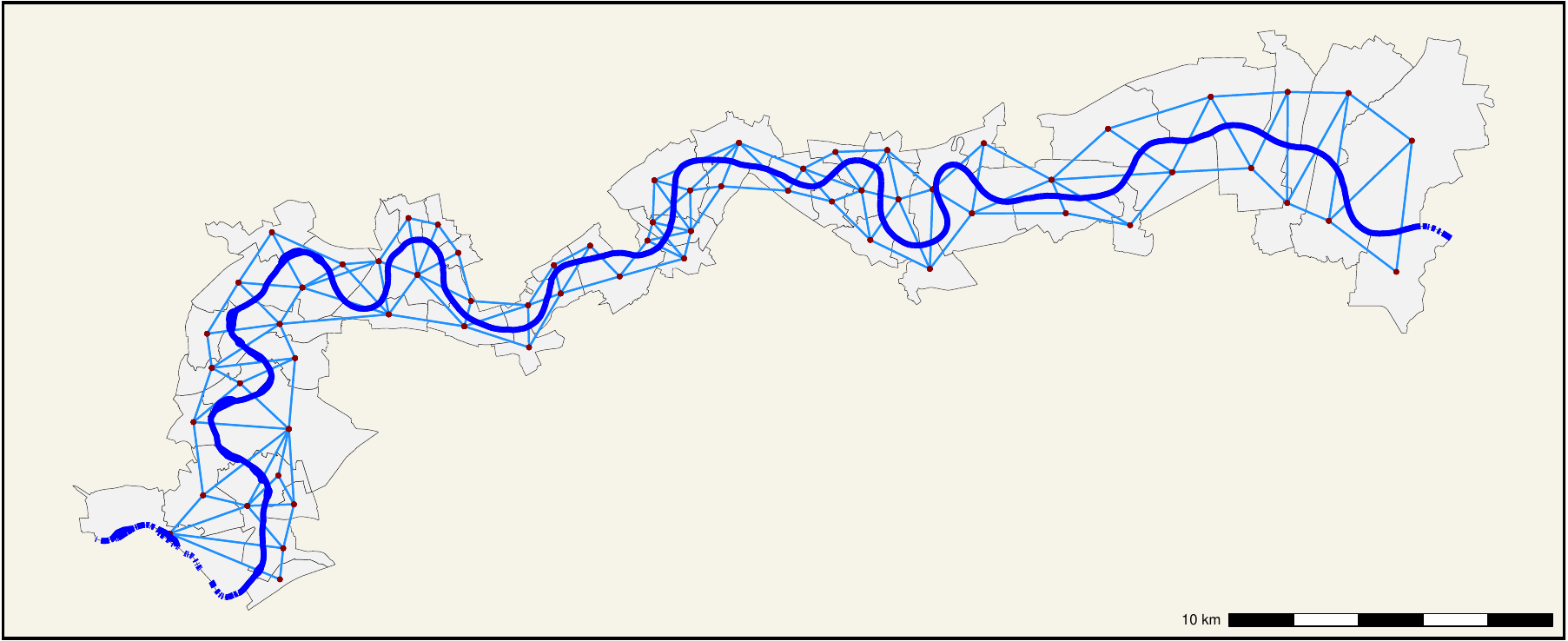} 

}

\caption{Riverside boroughs of Greater London. Queen contiguity disregarding river. }\label{fig:lon-contig2}
\end{figure}

\texttt{sfislands} provides convenient functions for this sort of situation. Let us start by restricting our wards and boroughs of interest to just those which are on either side of the river Thames. Figure \ref{fig:lon-contig2} shows the resultant contiguities when the river is ignored.

\begin{Shaded}
\begin{Highlighting}[]
\CommentTok{\# which boroughs are alongside the river}

\NormalTok{riverside }\OtherTok{\textless{}{-}}\NormalTok{ thames }\SpecialCharTok{|\textgreater{}} \FunctionTok{st\_intersects}\NormalTok{(london) }\SpecialCharTok{|\textgreater{}} \FunctionTok{unlist}\NormalTok{() }\SpecialCharTok{|\textgreater{}} \FunctionTok{unique}\NormalTok{()}

\CommentTok{\# only map these boroughs}

\FunctionTok{st\_bridges}\NormalTok{(london[riverside,],}\StringTok{"NAME"}\NormalTok{) }\SpecialCharTok{|\textgreater{}} 
  \FunctionTok{st\_quickmap\_nb}\NormalTok{(}\AttributeTok{linksize =} \FloatTok{0.5}\NormalTok{) }\SpecialCharTok{+}
  \FunctionTok{geom\_sf}\NormalTok{(}\AttributeTok{data=}\NormalTok{thames, }\AttributeTok{colour=}\StringTok{"blue"}\NormalTok{, }\AttributeTok{linewidth=}\FloatTok{1.5}\NormalTok{) }\SpecialCharTok{+} 
  \FunctionTok{annotation\_scale}\NormalTok{(}\AttributeTok{location=}\StringTok{"br"}\NormalTok{) }\SpecialCharTok{+}
  \FunctionTok{coord\_sf}\NormalTok{(}\AttributeTok{datum=}\ConstantTok{NA}\NormalTok{) }\SpecialCharTok{+} 
  \FunctionTok{theme}\NormalTok{(}\AttributeTok{panel.background =} \FunctionTok{element\_rect}\NormalTok{(}\AttributeTok{fill =} \StringTok{"\#F6F3E9"}\NormalTok{, }\AttributeTok{colour =} \StringTok{"black"}\NormalTok{, }
                                        \AttributeTok{linewidth=}\FloatTok{1.5}\NormalTok{))}
\end{Highlighting}
\end{Shaded}

\begin{figure}

{\centering \includegraphics[width=1\linewidth]{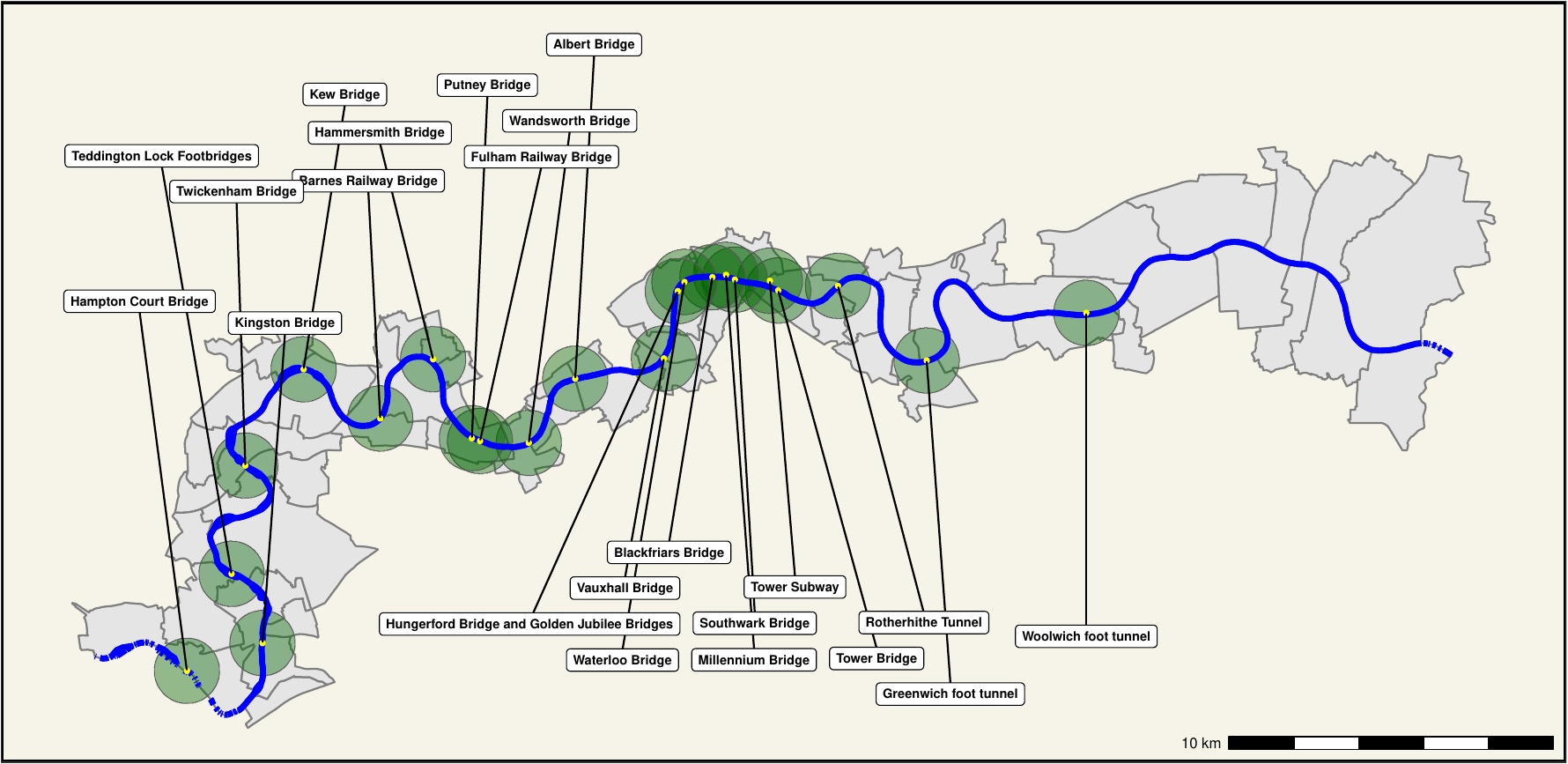} 

}

\caption{Riverside boroughs of Greater London. Road and pedestrian crossing and tunnels surrounded by 1 kilometre buffer shaded green. }\label{fig:lon-crossings}
\end{figure}

In order to take account of actual connectivity, we can add a layer showing the road and pedestrian bridges or tunnels. Details of these were sourced from the Wikipedia (2024) article titled ``\emph{List of crossings of the River Thames}''. In Figure \ref{fig:lon-crossings}, we have also drawn a 1 kilometre buffer around each crossing. This was chosen as an arbitrary measure of what might be considered a ``reasonable'' distance within which to consider opposing banks as being connected. The vast majority of units on opposing banks have access to a river crossing within this threshold and thus should be considered as neighbours. Only the extreme eastern units and one to the south west should not have a connection across the river according to this criterion.

\begin{figure}

{\centering \includegraphics[width=1\linewidth]{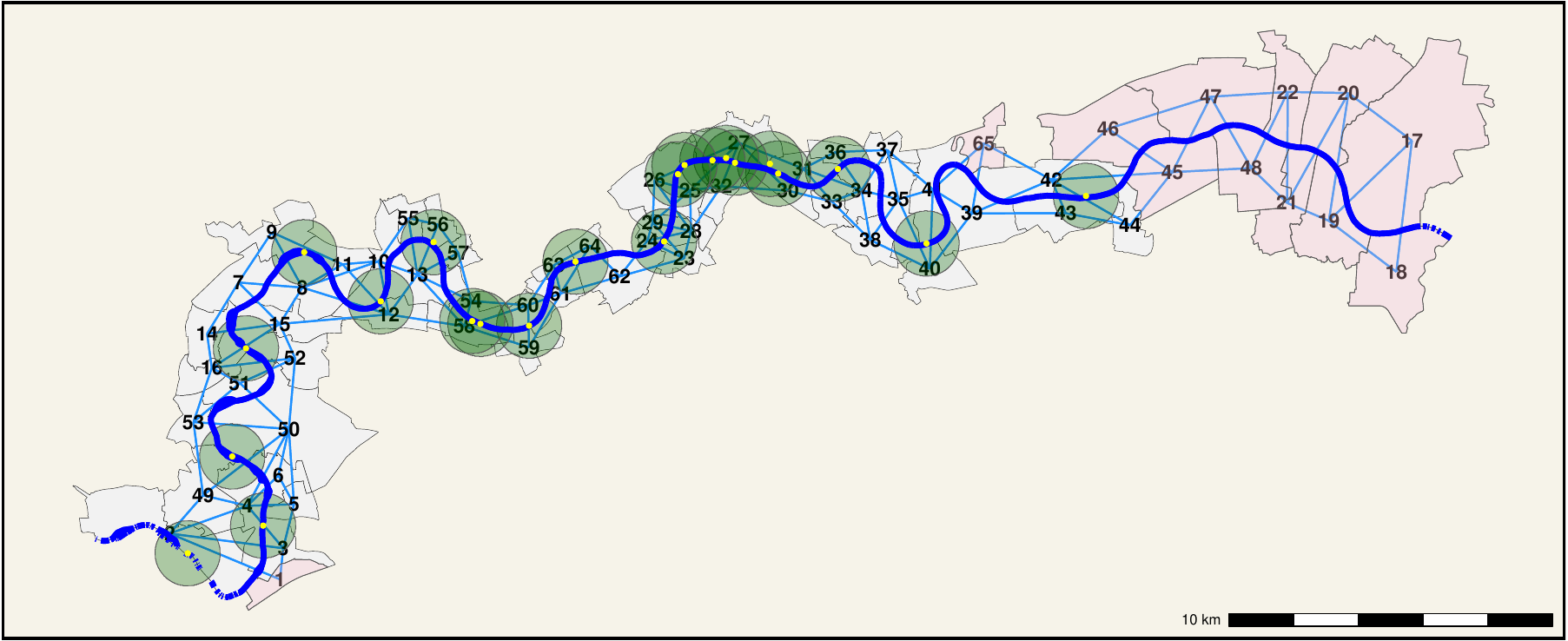} 

}

\caption{Riverside boroughs of Greater London. Index number for each borough shown at centroid. Boroughs which are not within 1 kilometre of a crossing are shaded pink. }\label{fig:lon-contig3}
\end{figure}

In order the identify the changes we wish to make, we use the
\texttt{nodes\ =\ "numeric"} argument in \texttt{st\_quickmap\_nb()}. Now we can identify
each unit by its position in the contiguity structure. Here we have
shaded in pink the units which are not within 1 kilometre of a river
crossing (see Figure \ref{fig:lon-contig3}).

\begin{Shaded}
\begin{Highlighting}[]
\CommentTok{\# with \textquotesingle{}nodes = "numeric"\textquotesingle{}}

\FunctionTok{st\_bridges}\NormalTok{(london[riverside,],}\StringTok{"NAME"}\NormalTok{) }\SpecialCharTok{|\textgreater{}} 
  \FunctionTok{st\_quickmap\_nb}\NormalTok{(}\AttributeTok{nodes =} \StringTok{"numeric"}\NormalTok{, }\AttributeTok{numericsize =} \DecValTok{4}\NormalTok{, }\AttributeTok{linksize =} \FloatTok{0.5}\NormalTok{) }\SpecialCharTok{+}
  \FunctionTok{geom\_sf}\NormalTok{(}\AttributeTok{data=}\NormalTok{no\_touch\_buffer, }\AttributeTok{fill=}\StringTok{"pink"}\NormalTok{, }\AttributeTok{alpha=}\FloatTok{0.3}\NormalTok{) }\SpecialCharTok{+} 
  \FunctionTok{geom\_sf}\NormalTok{(}\AttributeTok{data=}\NormalTok{crossings\_roadped }\SpecialCharTok{|\textgreater{}} \FunctionTok{st\_buffer}\NormalTok{(}\DecValTok{1000}\NormalTok{), }
          \AttributeTok{fill=}\StringTok{"darkgreen"}\NormalTok{, }\AttributeTok{alpha=}\FloatTok{0.3}\NormalTok{) }\SpecialCharTok{+}
  \FunctionTok{geom\_sf}\NormalTok{(}\AttributeTok{data=}\NormalTok{thames, }\AttributeTok{colour=}\StringTok{"blue"}\NormalTok{, }\AttributeTok{linewidth=}\FloatTok{1.5}\NormalTok{) }\SpecialCharTok{+} 
  \FunctionTok{geom\_sf}\NormalTok{(}\AttributeTok{data=}\NormalTok{crossings\_roadped, }\AttributeTok{size=}\DecValTok{1}\NormalTok{, }\AttributeTok{colour=}\StringTok{"yellow"}\NormalTok{) }\SpecialCharTok{+} 
  \FunctionTok{annotation\_scale}\NormalTok{(}\AttributeTok{location=}\StringTok{"br"}\NormalTok{) }\SpecialCharTok{+}
  \FunctionTok{coord\_sf}\NormalTok{(}\AttributeTok{datum=}\ConstantTok{NA}\NormalTok{) }\SpecialCharTok{+} 
  \FunctionTok{theme}\NormalTok{(}\AttributeTok{panel.background =} \FunctionTok{element\_rect}\NormalTok{(}\AttributeTok{fill =} \StringTok{"\#F6F3E9"}\NormalTok{, }\AttributeTok{colour =} \StringTok{"black"}\NormalTok{, }
                                        \AttributeTok{linewidth=}\FloatTok{1.5}\NormalTok{))}
\end{Highlighting}
\end{Shaded}

\begin{figure}

{\centering \includegraphics[width=1\linewidth]{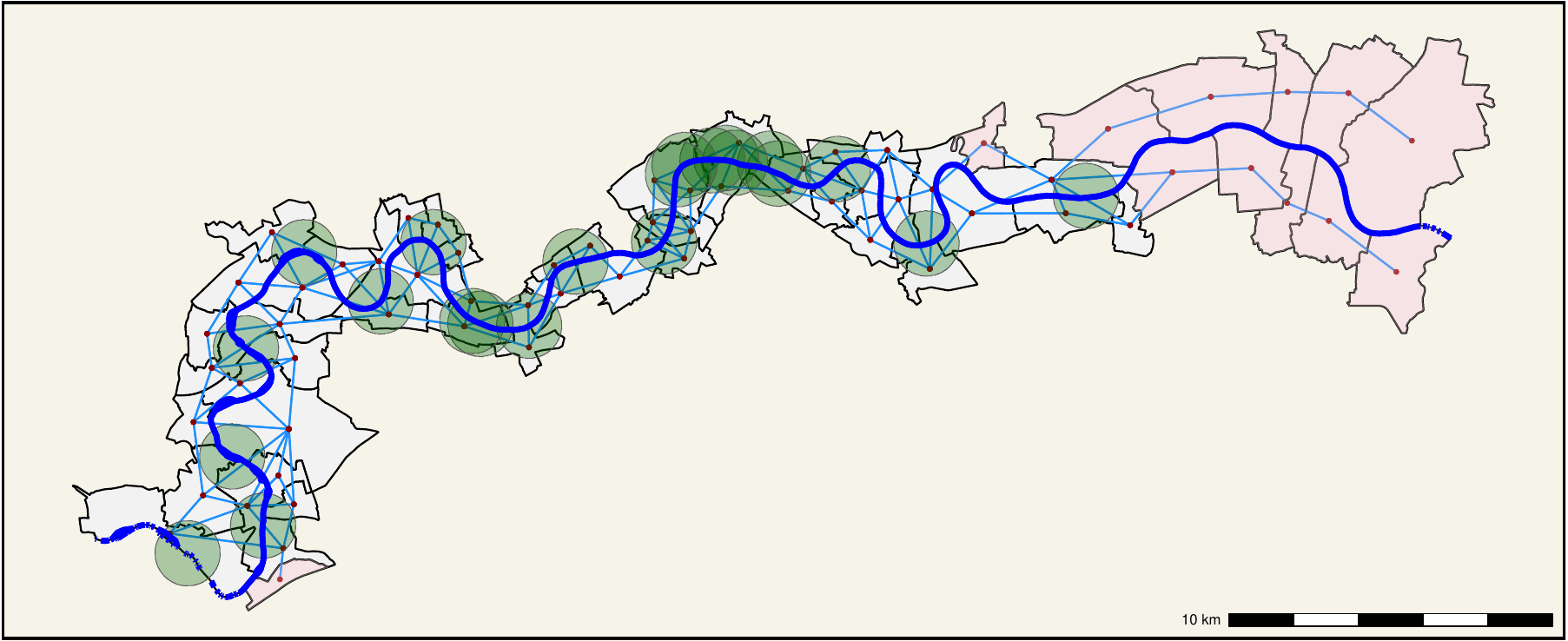} 

}

\caption{Riverside boroughs of Greater London. Contiguities across the river have been cut for the pink boroughs. }\label{fig:lon-contig4}
\end{figure}

This allows us to easily cut the ties across the river for these units
manually by using the function \texttt{st\_manual\_cut\_nb()}. \footnote{While we are using the index of the units in this example, the function also
  accepts names as arguments which may be more convenient in some circumstances.} Having made these adjustments,
\texttt{st\_quickmap\_nb()} now shows a connectivity structure (Figure \ref{fig:lon-contig4}) which reflects our hypothesis of how influence should extend across the river in the presence or absence of crossings.

This example shows that the pre-functions of \texttt{sfislands} have uses for situations which do not involve islands. They can be used to apply domain knowledge to easily design the most appropriate neighbourhood structure.

\begin{Shaded}
\begin{Highlighting}[]
\CommentTok{\# manually cut the links where there is no crossing}

\FunctionTok{st\_bridges}\NormalTok{(london[riverside,], }\StringTok{"NAME"}\NormalTok{) }\SpecialCharTok{|\textgreater{}} 
  \FunctionTok{st\_manual\_cut\_nb}\NormalTok{(}\DecValTok{18}\NormalTok{,}\DecValTok{17}\NormalTok{) }\SpecialCharTok{|\textgreater{}} 
  \FunctionTok{st\_manual\_cut\_nb}\NormalTok{(}\DecValTok{19}\NormalTok{,}\DecValTok{17}\NormalTok{) }\SpecialCharTok{|\textgreater{}} 
  \FunctionTok{st\_manual\_cut\_nb}\NormalTok{(}\DecValTok{19}\NormalTok{,}\DecValTok{20}\NormalTok{) }\SpecialCharTok{|\textgreater{}} 
  \FunctionTok{st\_manual\_cut\_nb}\NormalTok{(}\DecValTok{20}\NormalTok{,}\DecValTok{21}\NormalTok{) }\SpecialCharTok{|\textgreater{}} 
  \FunctionTok{st\_manual\_cut\_nb}\NormalTok{(}\DecValTok{21}\NormalTok{,}\DecValTok{22}\NormalTok{) }\SpecialCharTok{|\textgreater{}} 
  \FunctionTok{st\_manual\_cut\_nb}\NormalTok{(}\DecValTok{22}\NormalTok{,}\DecValTok{48}\NormalTok{) }\SpecialCharTok{|\textgreater{}} 
  \FunctionTok{st\_manual\_cut\_nb}\NormalTok{(}\DecValTok{47}\NormalTok{,}\DecValTok{48}\NormalTok{) }\SpecialCharTok{|\textgreater{}} 
  \FunctionTok{st\_manual\_cut\_nb}\NormalTok{(}\DecValTok{45}\NormalTok{,}\DecValTok{46}\NormalTok{) }\SpecialCharTok{|\textgreater{}} 
  \FunctionTok{st\_manual\_cut\_nb}\NormalTok{(}\DecValTok{45}\NormalTok{,}\DecValTok{47}\NormalTok{) }\SpecialCharTok{|\textgreater{}}  
  \FunctionTok{st\_manual\_cut\_nb}\NormalTok{(}\DecValTok{39}\NormalTok{,}\DecValTok{65}\NormalTok{) }\SpecialCharTok{|\textgreater{}} 
  \FunctionTok{st\_manual\_cut\_nb}\NormalTok{(}\DecValTok{1}\NormalTok{,}\DecValTok{2}\NormalTok{) }\SpecialCharTok{|\textgreater{}} 
  \FunctionTok{st\_quickmap\_nb}\NormalTok{(}\AttributeTok{bordercol =} \StringTok{"black"}\NormalTok{, }\AttributeTok{bordersize =} \FloatTok{0.5}\NormalTok{, }\AttributeTok{linksize =} \FloatTok{0.5}\NormalTok{) }\SpecialCharTok{+}
  \FunctionTok{geom\_sf}\NormalTok{(}\AttributeTok{data=}\NormalTok{no\_touch\_buffer, }\AttributeTok{fill =} \StringTok{"pink"}\NormalTok{, }\AttributeTok{alpha =} \FloatTok{0.3}\NormalTok{) }\SpecialCharTok{+} 
  \FunctionTok{geom\_sf}\NormalTok{(}\AttributeTok{data=}\NormalTok{crossings\_roadped }\SpecialCharTok{|\textgreater{}} \FunctionTok{st\_buffer}\NormalTok{(}\DecValTok{1000}\NormalTok{), }
          \AttributeTok{fill=} \StringTok{"darkgreen"}\NormalTok{, }\AttributeTok{alpha =} \FloatTok{0.3}\NormalTok{) }\SpecialCharTok{+}
  \FunctionTok{geom\_sf}\NormalTok{(}\AttributeTok{data=}\NormalTok{thames, }\AttributeTok{colour =} \StringTok{"blue"}\NormalTok{, }\AttributeTok{linewidth =} \FloatTok{1.5}\NormalTok{) }\SpecialCharTok{+}
  \FunctionTok{annotation\_scale}\NormalTok{(}\AttributeTok{location =} \StringTok{"br"}\NormalTok{) }\SpecialCharTok{+}
  \FunctionTok{coord\_sf}\NormalTok{(}\AttributeTok{datum=}\ConstantTok{NA}\NormalTok{) }\SpecialCharTok{+} 
  \FunctionTok{theme}\NormalTok{(}\AttributeTok{panel.background =} \FunctionTok{element\_rect}\NormalTok{(}\AttributeTok{fill =} \StringTok{"\#F6F3E9"}\NormalTok{, }\AttributeTok{colour =} \StringTok{"black"}\NormalTok{, }
                                        \AttributeTok{linewidth =} \FloatTok{1.5}\NormalTok{))}
\end{Highlighting}
\end{Shaded}

\section{Liverpool (example 3)}\label{liverpool-example-3}

In this final example, there are no islands and no river issues. We saw in the Indonesia example how \texttt{sfislands} can extract estimates from the \texttt{mgcv} argument \texttt{bs="mrf"}. It can also deal with random effects as defined by \texttt{bs="re"} in \texttt{mgcv} models. This means it is possible to create quite complex models featuring both a traditional multi-level structure and conditionally correlated components. Here, we will construct a hierarchical model in this way which also features an ICAR component at the lowest level of the tree. We will compare two different types of contiguity which could be applied to the ICAR component:

\begin{itemize}
\tightlist
\item
  \texttt{st\_bridges()} from \texttt{sfislands}, which implements first-order queen contiguity in the same way as \texttt{sfdep:st\_contiguity()} for situations such as this where there are no islands, and
\item
  \texttt{st\_dist\_band()} from \texttt{sfdep}, which connects units whose centroids lie within a chosen distance threshold.
\end{itemize}

We will show how their outcomes can be quickly mapped to give an immediate sense of the differences implied by each model's structure.

\begin{figure}

{\centering \includegraphics[width=1\linewidth]{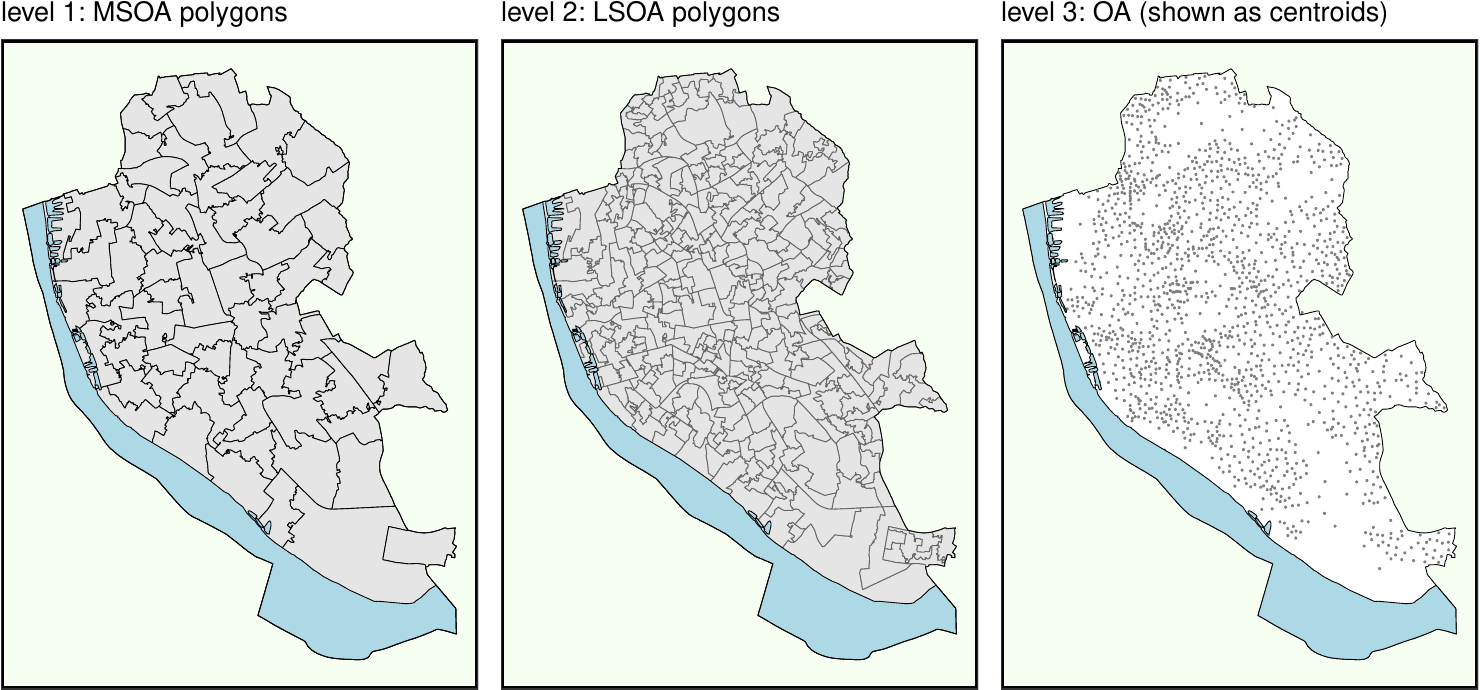} 

}

\caption{Nested levels of administrative divisions in Liverpool. }\label{fig:liv-nested}
\end{figure}

This example is based on data from the \href{https://www.nomisweb.co.uk/home/census2001.asp}{Office of National Statistics} as presented by Rowe and Arribas-Bel (2024) in their demonstration of techniques for hierarchical spatial regression. They use the different levels of nested administrative boundaries in Liverpool for their demonstration. These are described in Table \ref{tab:livadmin-latex}, and are mapped separately in Figure \ref{fig:liv-nested}.

\begin{figure}

{\centering \includegraphics[width=0.7\linewidth]{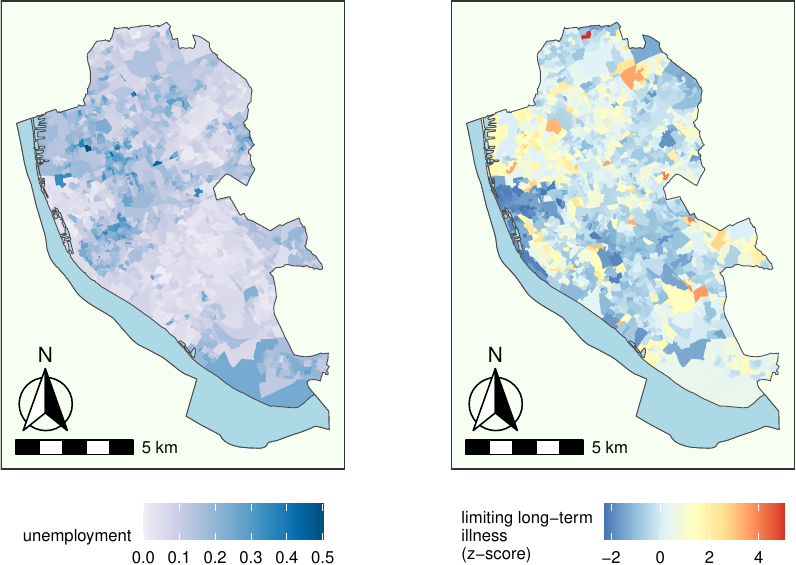} 

}

\caption{Unemployment and limiting long-term illness rates in Liverpool, with extent of Mersey estuary in light blue to the south west. }\label{fig:liv-vars}
\end{figure}

Now suppose we wish to model unemployment as a function of limiting long-term illness, taking into account this hierarchical structure. Limiting long-term illness (LLTI) refers to a health problem or disability which limits a person's day-to-day activities (Barnett 2001). We wish to incorporate a fixed intercept and coefficient for LLTI, with additional random intercepts and slopes at different levels. The raw distribution of these variables across the output areas (OAs) of Liverpool are shown in Figure
\ref{fig:liv-vars}.

If \(y_{ijk}\) and \(x_{ijk}\) are respectively the levels of unemployment and LLTI in each of \(i\) OAs within \(j\) lower level super output areas (LSOAs), which in turn are within \(k\) middle level super output areas (MSOAs), then
\begin{equation}
y_{ijk} = \beta_{0jk} + \beta_{1jk}x_{ijk} + \gamma_{ijk} + \epsilon_{ijk}
\label{eq:eq9}
\end{equation}
\begin{equation}
\beta_{0jk} = \beta_{0k} + u_{0jk}
\label{eq:eq10}
\end{equation}
\begin{equation}
\beta_{1jk} = \beta_{1k} + u_{1jk}
\label{eq:eq11}
\end{equation}
\begin{equation}
\beta_{0k} = \beta_0 + w_{0k}
\label{eq:eq12}
\end{equation}
\begin{equation}
\beta_{1k} = \beta_1 + w_{1k}
\label{eq:eq13}
\end{equation}
\begin{equation}
\epsilon_{ijk} \sim N(0,\sigma_{\epsilon}^2)
\label{eq:eq14}
\end{equation}
\begin{equation}
u_{jk} \sim N(0,\sigma_{u}^2)
\label{eq:eq15}
\end{equation}
\begin{equation}
 w_{k} \sim N(0,\sigma_{w}^2)
\label{eq:eq16}
\end{equation}
where the \(\gamma_{ijk}\)s comprise a vector of ICAR components for each OA, similar in form to the ICAR component for the provinces in the earlier Indonesia model. The vector has a multivariate normal distribution with mean 0 and precision related to the contiguity structure as before.

\begin{figure}

{\centering \includegraphics[width=0.6\linewidth]{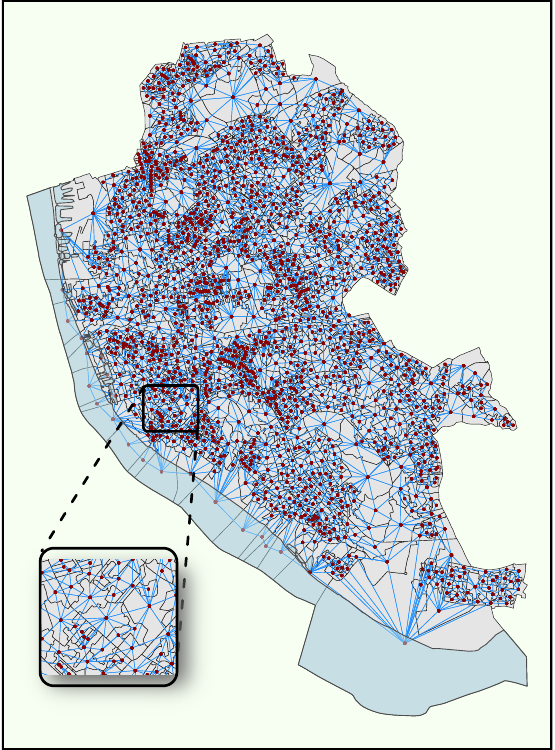} 

}

\caption{First order queen contiguity of output areas (OAs) in Liverpool with magnified inset. }\label{fig:liv-contigs}
\end{figure}

There are no islands or discontiguities in this geometry of Liverpool so using \texttt{st\_bridges()} will give the same structure as would \texttt{sfdep::st\_contiguity()} (see Figure \ref{fig:liv-contigs}, with a magnified section for visibility purposes, created using \texttt{ggmagnify} (Hugh-Jones 2024)).

\begin{table}
\centering
\caption{\label{tab:livadmin-latex}Three nested levels of administrative divisions in Liverpool.}
\centering
\fontsize{8}{10}\selectfont
\begin{tabular}[t]{l|l|>{\raggedright\arraybackslash}p{7cm}}
\hline
\textbf{code} & \textbf{area type} & \textbf{description}\\
\hline
OA & output areas & lowest level of geographical area for census statistics, usually containing 100 - 625 persons\\
\hline
LSOA & lower layer super output areas & usually 4 or 5 OAs\\
\hline
MSOA & middle layer super output areas & usually 4 or 5 LSOAs\\
\hline
\end{tabular}
\end{table}

\begin{Shaded}
\begin{Highlighting}[]
\CommentTok{\# show st\_bridges() connection with a magnified section}

\NormalTok{from }\OtherTok{\textless{}{-}} \FunctionTok{c}\NormalTok{(}\AttributeTok{xmin =} \SpecialCharTok{{-}}\FloatTok{2.97}\NormalTok{, }\AttributeTok{xmax =} \SpecialCharTok{{-}}\FloatTok{2.95}\NormalTok{, }\AttributeTok{ymin =} \FloatTok{53.38}\NormalTok{, }\AttributeTok{ymax =} \FloatTok{53.39}\NormalTok{)}
\NormalTok{to }\OtherTok{\textless{}{-}} \FunctionTok{c}\NormalTok{(}\SpecialCharTok{{-}}\FloatTok{3.01}\NormalTok{,}\SpecialCharTok{{-}}\FloatTok{2.96}\NormalTok{,}\FloatTok{53.32}\NormalTok{,}\FloatTok{53.35}\NormalTok{)}

\FunctionTok{st\_bridges}\NormalTok{(liverpool, }\StringTok{"oa\_cd"}\NormalTok{) }\SpecialCharTok{|\textgreater{}} 
  \FunctionTok{st\_quickmap\_nb}\NormalTok{(}\AttributeTok{pointsize =} \FloatTok{0.1}\NormalTok{,}
                 \AttributeTok{linksize =} \FloatTok{0.1}\NormalTok{,}
                 \AttributeTok{fillcol =} \StringTok{"gray90"}\NormalTok{) }\SpecialCharTok{+}  
  \FunctionTok{geom\_sf}\NormalTok{(}\AttributeTok{data=}\NormalTok{mersey, }\AttributeTok{fill=}\StringTok{"lightblue"}\NormalTok{, }\AttributeTok{colour=}\StringTok{"gray30"}\NormalTok{, }
          \AttributeTok{linewidth=}\FloatTok{0.2}\NormalTok{, }\AttributeTok{alpha=}\FloatTok{0.5}\NormalTok{) }\SpecialCharTok{+} 
  \FunctionTok{geom\_magnify}\NormalTok{(}\AttributeTok{data=}\NormalTok{liverpool, }
               \AttributeTok{from =}\NormalTok{ from, }\AttributeTok{to =}\NormalTok{ to, }
               \AttributeTok{corners =} \FloatTok{0.1}\NormalTok{, }\AttributeTok{shadow =} \ConstantTok{TRUE}\NormalTok{, }\AttributeTok{linewidth =} \FloatTok{0.6}\NormalTok{) }\SpecialCharTok{+} 
  \FunctionTok{theme}\NormalTok{(}\AttributeTok{panel.background =} \FunctionTok{element\_rect}\NormalTok{(}\AttributeTok{fill =} \StringTok{"\#F7FFF2"}\NormalTok{,}
                                        \AttributeTok{colour =} \StringTok{"black"}\NormalTok{, }\AttributeTok{linewidth=}\DecValTok{1}\NormalTok{))}
\end{Highlighting}
\end{Shaded}

\begin{figure}

{\centering \includegraphics[width=1\linewidth]{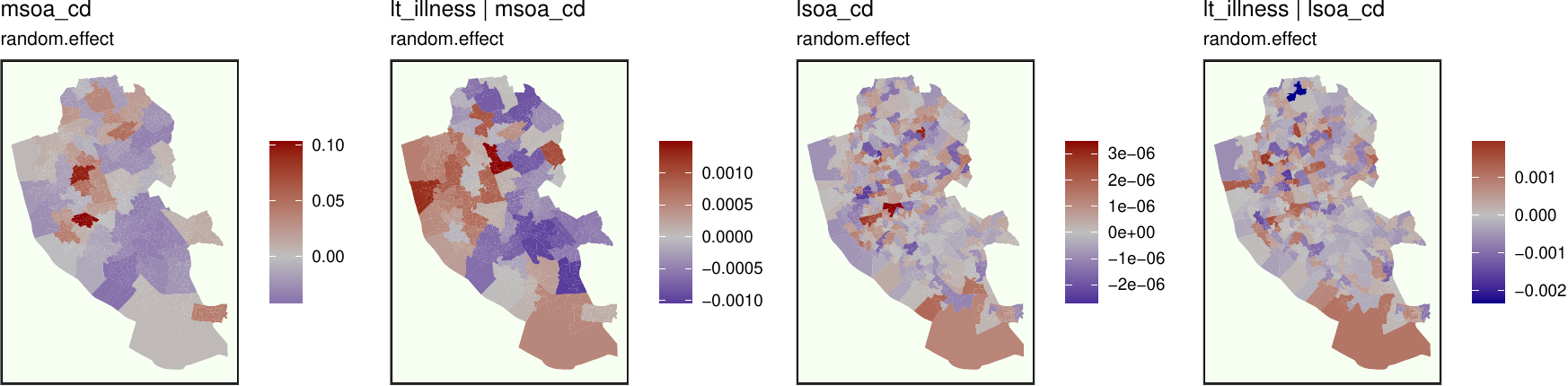} 

}

\caption{Random intercept and slope estimates from hierarchical model. Middle and lower layer super output areas as levels. }\label{fig:liv-est1}
\end{figure}

\begin{figure}

{\centering \includegraphics[width=1\linewidth]{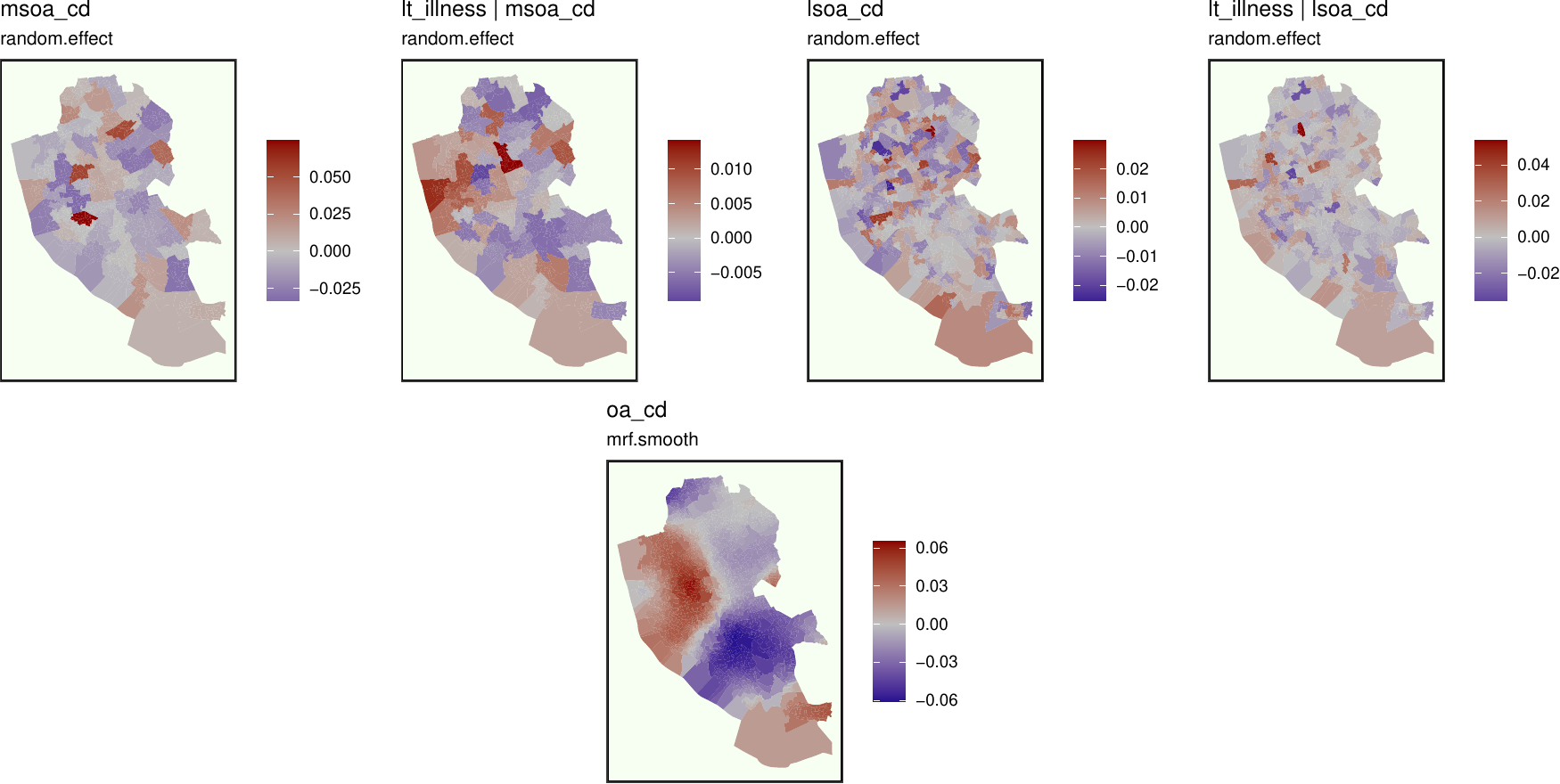} 

}

\caption{Random intercept and slope estimates from hierarchical model. Middle and lower layer super output areas are the top two levels respectively (MSOA \& LSOA). Output areas (OA) as lowest level Markov random field smooth using first-degree queen contiguity from \texttt{st\_bridges()}.}\label{fig:liv-est2}
\end{figure}

To demonstrate how \texttt{sfislands} can be used to quickly compare the output of models, we show the output of three similar models. The first of these is a hierarchical model as described above, but with no ICAR component at the lowest level. Its spatially varying coefficients are shown in Figure \ref{fig:liv-est1}. Secondly, we fit the model described above using \texttt{st\_bridges()} to generate first-order queen contiguities as the structure for the neighbourhood relationships. The coefficients of this model are shown in Figure \ref{fig:liv-est2}.

\begin{figure}

{\centering \includegraphics[width=1\linewidth]{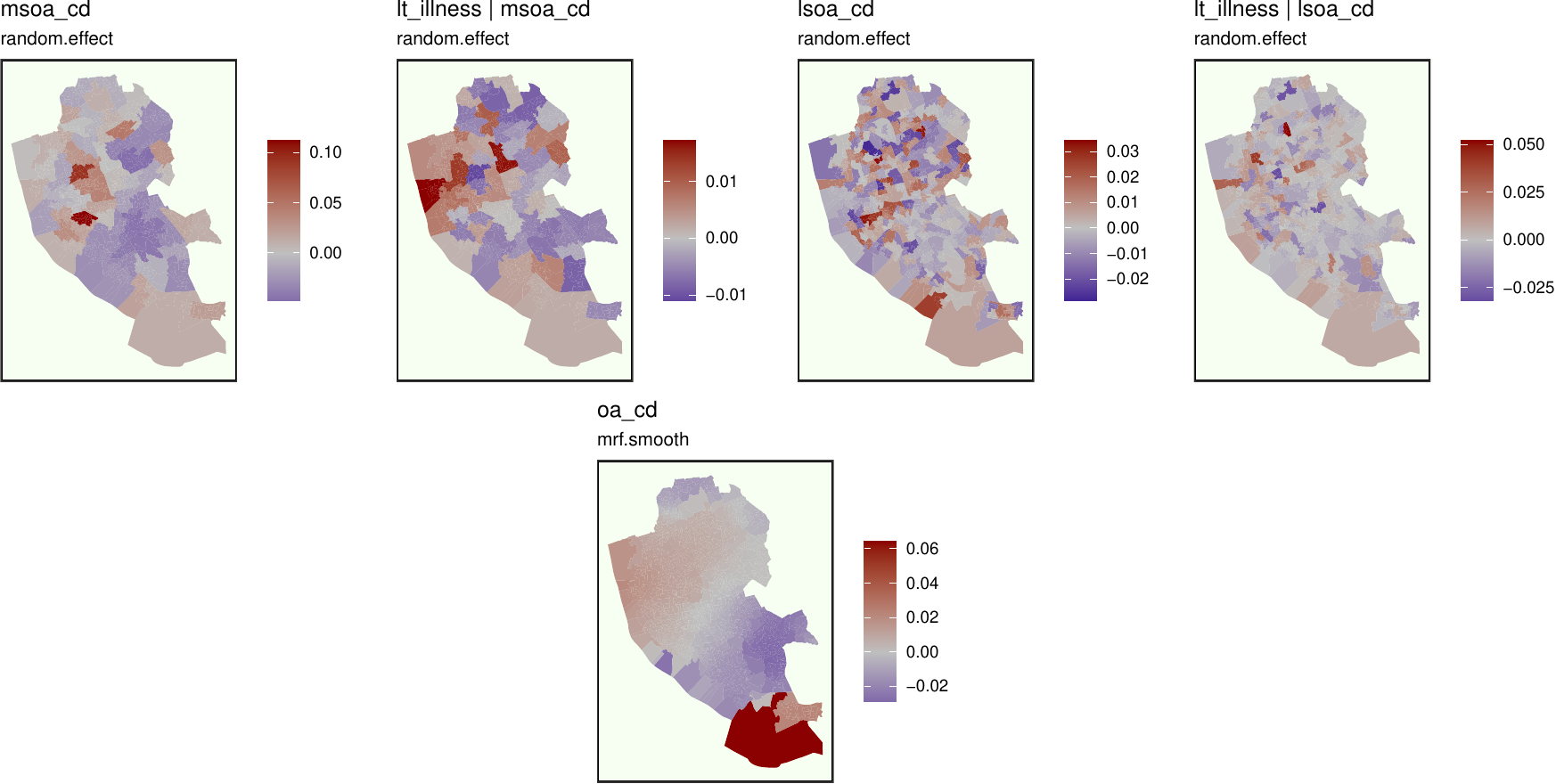} 

}

\caption{Random intercept and slope estimates from hierarchical model. Middle and lower layer super output areas are the top two levels respectively (MSOA \& LSOA). Output areas (OA) as lowest level Markov random field smooth using \texttt{sfdep::st\_dist\_band()} with radius of 700 metres.}\label{fig:liv-est3}
\end{figure}

Finally, we can fit a similar model but this time using a within-distance criterion of 700m for neighbourhood status. All units within this distance of each other are considered neighbours. This was generated with the \texttt{sfdep:st\_dist\_band()} function. These coefficients are shown in Figure \ref{fig:liv-est3}.

This shows how \texttt{sfislands} can quickly give an intuitive representation of how a set of similar models are operating. As a measure of the relative strengths of these three structures, we could compare their AICs (see Table \ref{tab:aic-latex}). These suggest that, in this scenario, inclusion of the ICAR component is appropriate, and queen contiguity is a more suitable choice than the distance-band condition for defining the neighbourhood structure.

\begin{table}
\centering
\caption{\label{tab:aic-latex}Comparison of models by AIC.}
\centering
\fontsize{8}{10}\selectfont
\begin{tabular}[t]{l|r|r}
\hline
\textbf{model} & \textbf{AIC} & \textbf{relative AIC}\\
\hline
hierarchical model, ICAR based on st\_bridges() & -5126.948 & -1.000\\
\hline
hierarchical model, ICAR based on sfdep::st\_dist\_band() & -5107.790 & -0.996\\
\hline
hierarchical model, no ICAR & -5082.684 & -0.991\\
\hline
\end{tabular}
\end{table}

\section{Summary}\label{summary}

These examples have shown the varying scenarios in which \texttt{sfislands} can be useful. It aims to contribute to spatial modelling by making an awkward area less awkward. Rather than having a default attitude of ignoring islands when building neighbourhood structures based on contiguity, it hopes to encourage at least an examination of whether or not it is appropriate for them to be included. Even when no islands are present, it provides a simple procedure for tailoring a neighbourhood structure to match a given hypothesis. It also provides helper functions to use these structures in spatial regression models, notably those built with \texttt{mgcv}, which streamline the human effort necessary to examine the estimates. In future, compatibility with other modelling packages can be added to broaden the package's capabilities.

\section{Acknowledgements}\label{acknowledgements}

This publication has emanated from research conducted with the financial support of Science Foundation Ireland under Grant number 18/CRT/6049.

\phantomsection\label{refs}

\bibliographystyle{unsrt}
\bibliography{references.bib}

\begin{CSLReferences}{1}{0}
\bibitem[\citeproctext]{ref-Andrfout2022}
Andréfouët, Serge, Mégane Paul, and A. Riza Farhan. 2022. {``Indonesia's 13558 Islands: A New Census from Space and a First Step Towards a One Map for Small Islands Policy.''} \emph{Marine Policy} 135 (January): 104848. \url{https://doi.org/10.1016/j.marpol.2021.104848}.

\bibitem[\citeproctext]{ref-r-inla}
Bakka, Haakon, Håvard Rue, Geir-Arne Fuglstad, Andrea Riebler, David Bolin, Elias Krainski, Daniel Simpson, and Finn Lindgren. 2018. {``Spatial Modelling with r-INLA: A Review.''} arXiv. \url{https://doi.org/10.48550/ARXIV.1802.06350}.

\bibitem[\citeproctext]{ref-Barnett2001}
Barnett, S. 2001. {``A Multilevel Analysis of the Effects of Rurality and Social Deprivation on Premature Limiting Long Term Illness.''} \emph{Journal of Epidemiology \& Community Health} 55 (1): 44--51. \url{https://doi.org/10.1136/jech.55.1.44}.

\bibitem[\citeproctext]{ref-lme4}
Bates, Douglas, Martin Mächler, Ben Bolker, and Steve Walker. 2015. {``Fitting Linear Mixed-Effects Models Using {lme4}.''} \emph{Journal of Statistical Software} 67 (1): 1--48. \url{https://doi.org/10.18637/jss.v067.i01}.

\bibitem[\citeproctext]{ref-besag}
Besag, Julian. 1974. {``Spatial Interaction and the Statistical Analysis of Lattice Systems.''} \emph{Journal of the Royal Statistical Society: Series B (Methodological)} 36 (2): 192--225. \url{https://doi.org/10.1111/j.2517-6161.1974.tb00999.x}.

\bibitem[\citeproctext]{ref-bivandportnov}
Bivand, Roger S., and Boris A. Portnov. 2004. {``{Exploring Spatial Data Analysis Techniques Using R: The Case of Observations with No Neighbors}.''} In \emph{{Advances in Spatial Econometrics}}, edited by Luc Anselin, Raymond J. G. M. Florax, and Sergio J. Rey, 121--42. Advances in Spatial Science. Springer. \url{https://doi.org/10.1007/978-3-662-05617-2}.

\bibitem[\citeproctext]{ref-spdep}
Bivand, Roger, L Anselin, O Berke, A Bernat, M Carvalho, Y Chun, CF Dormann, et al. 2011. {``Spdep: Spatial Dependence: Weighting Schemes, Statistics and Models.''} \emph{R Package Version 0.5-31, URL Http://CRAN. R-Project. Org/Package= Spdep}.

\bibitem[\citeproctext]{ref-BrizRedn2021}
Briz-Redón, Álvaro, Adina Iftimi, Juan Francisco Correcher, Jose De Andrés, Manuel Lozano, and Carolina Romero-García. 2021. {``A Comparison of Multiple Neighborhood Matrix Specifications for Spatio-Temporal Model Fitting: A Case Study on COVID-19 Data.''} \emph{Stochastic Environmental Research and Risk Assessment} 36 (1): 271--82. \url{https://doi.org/10.1007/s00477-021-02077-y}.

\bibitem[\citeproctext]{ref-brms}
Bürkner, Paul-Christian. 2017. {``{brms}: An {R} Package for {Bayesian} Multilevel Models Using {Stan}.''} \emph{Journal of Statistical Software} 80 (1): 1--28. \url{https://doi.org/10.18637/jss.v080.i01}.

\bibitem[\citeproctext]{ref-Duncan2017}
Duncan, Earl W., Nicole M. White, and Kerrie Mengersen. 2017. {``Spatial Smoothing in Bayesian Models: A Comparison of Weights Matrix Specifications and Their Impact on Inference.''} \emph{International Journal of Health Geographics} 16 (1). \url{https://doi.org/10.1186/s12942-017-0120-x}.

\bibitem[\citeproctext]{ref-Earnest2007}
Earnest, Arul, Geoff Morgan, Kerrie Mengersen, Louise Ryan, Richard Summerhayes, and John Beard. 2007. {``Evaluating the Effect of Neighbourhood Weight Matrices on Smoothing Properties of Conditional Autoregressive (CAR) Models.''} \emph{International Journal of Health Geographics} 6 (1): 54. \url{https://doi.org/10.1186/1476-072x-6-54}.

\bibitem[\citeproctext]{ref-freni}
Freni-Sterrantino, Anna, Massimo Ventrucci, and Håvard Rue. 2018. {``A Note on Intrinsic Conditional Autoregressive Models for Disconnected Graphs.''} \emph{Spatial and Spatio-Temporal Epidemiology} 26: 25--34.

\bibitem[\citeproctext]{ref-sfislands}
Horan, Kevin, Katarina Domijan, and Chris Brunsdon. 2024. \emph{{sfislands}: Streamlines the Process of Fitting Areal Spatial Models}. \url{https://horankev.github.io/sfislands/}.

\bibitem[\citeproctext]{ref-ggmagnify}
Hugh-Jones, David. 2024. \emph{Ggmagnify: Create a Magnified Inset of Part of a "Ggplot" Object}. \url{https://github.com/hughjonesd/ggmagnify}.

\bibitem[\citeproctext]{ref-ggpubr}
Kassambara, Alboukadel. 2023. \emph{{ggpubr}: {``{ggplot2}''} Based Publication Ready Plots}. \url{https://CRAN.R-project.org/package=ggpubr}.

\bibitem[\citeproctext]{ref-sfdep}
Parry, Josiah. 2023. \emph{{sfdep}: Spatial Dependence for Simple Features}. \url{https://CRAN.R-project.org/package=sfdep}.

\bibitem[\citeproctext]{ref-sf}
Pebesma, Edzer. 2018. {``{Simple Features for R: Standardized Support for Spatial Vector Data}.''} \emph{{The R Journal}} 10 (1): 439--46. \url{https://doi.org/10.32614/RJ-2018-009}.

\bibitem[\citeproctext]{ref-nlme}
Pinheiro, José, Douglas Bates, and R Core Team. 2023. \emph{Nlme: Linear and Nonlinear Mixed Effects Models}. \url{https://CRAN.R-project.org/package=nlme}.

\bibitem[\citeproctext]{ref-broom}
Robinson, David, Alex Hayes, and Simon Couch. 2023. {``Broom: Convert Statistical Objects into Tidy Tibbles.''} \url{https://CRAN.R-project.org/package=broom}.

\bibitem[\citeproctext]{ref-liverpool-notes}
Rowe, Francisco, and Dani Arribas-Bel. 2024. {``{University of Liverpool, Spatial Modeling for Data Scientists, ENVS453, Course repository}.''} \url{https://gdsl-ul.github.io/san/}.

\bibitem[\citeproctext]{ref-rstan}
Stan Development Team. 2020. {``{RStan}: The {R} Interface to {Stan}.''} \url{http://mc-stan.org/}.

\bibitem[\citeproctext]{ref-tobler}
Tobler, W. R. 1970. {``A Computer Movie Simulating Urban Growth in the Detroit Region.''} \emph{Economic Geography} 46 (sup1): 234--40. \url{https://doi.org/10.2307/143141}.

\bibitem[\citeproctext]{ref-ggplot2}
Wickham, Hadley. 2016. \emph{Ggplot2: Elegant Graphics for Data Analysis}. Springer-Verlag New York. \url{https://ggplot2.tidyverse.org}.

\bibitem[\citeproctext]{ref-tidyverse}
Wickham, Hadley, Mara Averick, Jennifer Bryan, Winston Chang, Lucy D'Agostino McGowan, Romain François, Garrett Grolemund, et al. 2019. {``Welcome to the {tidyverse}.''} \emph{Journal of Open Source Software} 4 (43): 1686. \url{https://doi.org/10.21105/joss.01686}.

\bibitem[\citeproctext]{ref-wikicrossings}
Wikipedia. 2024. {``{List of crossings of the River Thames} --- {W}ikipedia{,} the Free Encyclopedia.''} \url{http://en.wikipedia.org/w/index.php?title=List/\%20of/\%20crossings/\%20of/\%20the/\%20River/\%20Thames&oldid=1184426738}.

\bibitem[\citeproctext]{ref-mgcv}
Wood, S. N. 2011. {``Fast Stable Restricted Maximum Likelihood and Marginal Likelihood Estimation of Semiparametric Generalized Linear Models''} 73: 3--36. \url{https://CRAN.R-project.org/web/packages/mgcv/index.html}.

\end{CSLReferences}

\end{document}